\documentclass[aps,prd,amsfonts,nofootinbib,onecolumn,floatfix,prd]{revtex4}

\usepackage{graphics,graphicx}
\usepackage{amsmath,amssymb}
\usepackage{bm,color}
\usepackage{dcolumn}

\def\be{\begin{equation}}
\def\ee{\end{equation}}
\def\ba{\begin{eqnarray}}
\def\ea{\end{eqnarray}}
\def\nn{\nonumber}

\def\ctt{C_l^{TT}}
\def\cee{C_l^{EE}}
\def\cte{C_l^{TE}}
\def\cpp{C_l^{\psi \psi}}
\def\ctp{C_l^{T \psi}}
\def\cgg{C_l^{gg}}
\def\ctg{C_l^{Tg}}
\def\cpg{C_l^{\psi g}}
\def\cep{C_l^{E \psi}}
\def\ceg{C_l^{E g}}

\def\hctt{\hat{C}_l^{TT}}
\def\bctt{\bar{C}_l^{TT}}
\def\hcee{\hat{C}_l^{EE}}
\def\bcee{\bar{C}_l^{EE}}
\def\hcte{\hat{C}_l^{TE}}
\def\bcte{\bar{C}_l^{TE}}
\def\hcpp{\hat{C}_l^{\psi \psi}}
\def\bcpp{\bar{C}_l^{\psi \psi}}
\def\hctp{\hat{C}_l^{T \psi}}
\def\bctp{\bar{C}_l^{T \psi}}
\def\hcgg{\hat{C}_l^{gg}}
\def\bcgg{\bar{C}_l^{gg}}
\def\hctg{\hat{C}_l^{Tg}}
\def\bctg{\bar{C}_l^{Tg}}
\def\hcpg{\hat{C}_l^{\psi g}}
\def\bcpg{\bar{C}_l^{\psi g}}

\begin{document}

\title{Constraints on primordial non-Gaussianity from Galaxy-CMB lensing cross-correlation}

\author{Yoshitaka Takeuchi$^1$, Kiyotomo Ichiki$^1$ and Takahiko Matsubara$^2$}
\affiliation{$^1$ Department of Physics, Nagoya
University, Chikusa-ku, Nagoya,  464-8602, Japan}
\affiliation{$^2$ Kobayashi-Maskawa Institute for the Origin of Particles and the Universe, Nagoya
University, Chikusa-ku, Nagoya,  464-8602, Japan}

\date{\today}

\begin{abstract}
  Recent studies have shown that the primordial non-Gaussianity
  affects clustering of dark matter halos through a scale-dependent
  bias and various constraints on the non-Gaussianity through this
  scale-dependent bias have been placed.  Here we introduce the
  cross-correlation between the CMB lensing potential and the galaxy
  angular distribution to effectively extract information about the
  bias from the galaxy distribution.  Then, we estimate the error of
  non-linear parameter, $f_{\rm NL}$, for the on-going CMB experiments
  and galaxy surveys, such as Planck and Hyper Suprime-Cam (HSC). We
  found that for the constraint on $f_{\rm NL}$ with Planck and HSC,
  the wide field galaxy survey is preferable to the deep one, and the
  expected error on $f_{\rm NL}$ can be as small as: $\Delta f_{\rm
    NL} \sim 20$ for $b_0 = 2$ and $\Delta f_{\rm NL} \sim 10$ for
  $b_0 = 4$, where $b_0$ is the linear bias parameter.  It is also
  found that future wide field galaxy survey could achieve $\Delta
  f_{\rm NL} \sim 5$ with CMB prior from Planck if one could observe
  highly biased objects at higher redshift ($z\sim 2$).
\end{abstract}



\maketitle

\section{Introduction}
\label{sec:Intro}
The cosmic microwave background (CMB) temperature anisotropy is 
a quite useful probe for cosmology. 
The contribution to the anisotropy is dominated by fluctuations 
at the last scattering surface. The CMB photon,
however, encounters the large-scale structure along the line of sight 
and some additional effects are imprinted on the temperature and
polarization as secondary anisotropies. The deflection of the CMB photon
due to gravitational potential produced by the large-scale structure
is one of them. The effect of the gravitational lensing on the CMB photon
through the large-scale structure is known as the CMB lensing
\cite{Lewis-06}. An on-going CMB observation by Planck \cite{Planck}
or various ground-based experiments are expected to detect this
signal, while the effect of the CMB lensing are imprinted on small
scales which the Wilkinson Microwave Anisotropy Probe (WMAP) satellite
could not resolve. The lensing effect reflects the late time
evolution of the universe at relatively low-redshifts. 
Therefore, the lensing information plays an important role
to determine the cosmological parameters, such as the neutrino mass,
the cosmological constant, the equation of state parameter of dark energy and
so on.

The large-scale structures are formed at relatively late time and they
become the source of the gravitational potential. They are correlated 
with the CMB temperature anisotropy through
the Integrated Sachs-Wolfe (ISW) effect, which generates the secondary
anisotropies due to the time variation of the potential \cite{Sachs-67}. 
The cross-correlation has an advantage for observations of ISW effect
whose signal is weak. Cross-correlations with complementary probes
are expected to provide additional information on top of their
respective auto-correlations. Similarly, we expect that lensing
potential should correlate with the large-scale structure and the
information from their cross-correlation may precisely determine the
cosmological parameters. \\

Recently, the deviations from Gaussian initial conditions
(primordial non-Gaussianity) are  intensively focused on and discussed. Inspection
of them offers an important window into the very early Universe
because non-standard models of inflation allow for a large
non-Gaussianity while standard single-field slow-roll models predict
the small deviations from Gaussianity. 
The most popular method to detect the primordial non-Gaussianity is to
measure higher-order correlations of CMB anisotropies and distributions
of galaxies, for example, the bispectrum or the three-point
correlation function of CMB \cite{Komatsu-01, Bartolo-04} or 
the large-scale structure bispectrum \cite{Scoccimarro-00, Verde-00,
  Scoccimarro-04, Sefusatti-07}.

Some studies have shown that the primordial non-Gaussianity affects
clustering of dark matter halos through a scale-dependent bias, both
by analytic calculations and by N-body simulations \cite{Afshordi-08,
  Carbone-08, Dalal-08, Pillepich-10, Slosar-08}.  By considering the
scale-dependent bias, one can constrain on the non-Gaussianity by the
power spectrum. Various constraints on the non-Gaussianity through the
scale-dependent bias have already been placed \cite{Slosar-08,
  Oguri-09, Carbone-10, Cunha-10, DeBernardis-10, Xia-10}. One may
expect, however, that a constraint on non-linear parameter $f_{\rm
  NL}$, which describes the primordial non-Gaussianity, is degenerated
with other cosmological parameters and has a large error.  Because of
such degeneracy, especially with the linear bias $b_0$, it is
important to combine unbiased observations which are sensitive to the
matter power spectrum in the near universe, like CMB lensing, shear
and so on.  Here we use the cross-correlation between the CMB lensing
potential and the large-scale galaxy distribution and estimate
expected errors of non-linear parameter $f_{\rm NL}$ for future CMB
experiments and galaxy surveys.  We expect that this cross-correlation
may play a role to break degeneracy and give us more stringent
constraints.

This paper is organized as follows. We review the scale-dependent bias
due to the non-Gaussianity in section \ref{sec:SBias} and the theory
behind the cross-correlation between CMB lensing potential and galaxy
distribution in section \ref{sec:APS}. In section \ref{sec:Modeling}
we describe the survey model for the galaxy distribution 
in the Hyper Suprime-Cam (HSC) survey, which is a fully funded imaging
survey at Subaru telescope. In section \ref{sec:Fisher} and section
\ref{sec:Forecasts} we explain the method of our analysis. Finally,
in section \ref{sec:Result} and \ref{sec:Summary} we discuss the
results and summarize our conclusions. 
Throughout this paper we assume a spatially flat universe for simplicity.

\section{Scale-Dependent Bias}
\label{sec:SBias}
Deviations from Gaussian initial conditions are commonly parameterized
in term of the dimensionless $f_{\rm NL}$ parameter and primordial
non-Gaussianity of the local-type is defined as \cite{Komatsu-01}
\begin{equation}
\Phi = \phi +
f_{\rm NL} (\phi^2 - \langle \phi^2 \rangle) , 
\end{equation}
where $\Phi$ denotes Bardeen's gauge-invariant potential and $\phi$
denotes a Gaussian random field. On subhorizon scale, $\Phi = - \Psi$,
where $\Psi$ denotes the usual Newtonian gravitational potential
related to density fluctuations via Poisson's equation. For example,
simple slow-role inflation gives a parameter $f_{\rm NL}$ of the order
of $10^{-2} - 1$ \cite{Salopek-90, Gangui-94, Falk-93}. On the other
hand, large values of $f_{\rm NL}$ can be expected in models of
multifield inflation, tachyonic preheating in hybrid inflation
\cite{Barnaby-06} or ghost inflation \cite{Arkani-Hamed-04}, for
instance.  Thus, the information about the inflation physics is
closely related to the parameter $f_{\rm NL}$.

Recent studies show that the effect of the primordial non-Gaussianity
of the local-type is seen in the clustering of halos through a
scale-dependent bias,
\begin{equation}
P_g(k)
= b_0^2 P(k) \rightarrow [b_0 + \Delta b(k)]^2 P(k) , 
\end{equation}
where $P_g (k)$ and $P (k)$ are the power spectrum of galaxy and
matter density fluctuations as a function of the wave number $k$,
respectively. $b_0$ is the Gaussian-case bias which relates the galaxy
density fluctuations with the matter density fluctuations, and $\Delta
b(k)$ represents the scale-dependence due to the non-Gaussianity
\cite{Afshordi-08, Carbone-08, Dalal-08, Pillepich-10, Slosar-08},
\begin{equation}
  \Delta b(k) = 
  \dfrac{3(b_0 -1) f_{\rm NL} \Omega_m H_0^2 \delta_c}{D(z) k^2 T(k)} ,  
\end{equation}
where $D(z)$ and $T(k)$ are the growth rate and the transfer function
for linear matter density fluctuations, respectively.  $\delta_c
\simeq 1.68$ is the threshold linear density contrast for a spherical
collapse of an overdensity region.  Primordial non-Gaussianity of the
local-type gives rise to a strong scale-dependent bias on large
scales, while the bias is roughly constant on large scales in the
Gaussian case.  However it is necessary to emphasize that the
constraint through the scale-dependent bias is sensitive only to the
local-type non-Gaussianity. For the constraints on the other
non-Gaussianity models we must consider higher-order correlation such
as bispectrum, trispectrum and so on.

\section{The Angular Power Spectrum}
\label{sec:APS}
The cross-correlations, for example, between CMB and galaxy, are well
known as providing additional information other than their respective
auto-correlation. In Ref.~\cite{Jeong-09}, they investigated the
cross-correlation between the shear of CMB lensing and halos. In this
paper, we introduce the cross-correlation between the CMB lensing and
galaxy angular distribution to estimate errors in constraining
cosmological parameters.

\subsection{Galaxy Distribution}
Probably the most obvious tracers of the large-scale density
field in the linear regime are luminous sources such as galaxies at
optical wavelengths and AGNs at x-rays and/or radio wavelengths.  The
projected density contrast of the tracers can be written as
\begin{equation}
  \delta_g (\hat{\bm n}) = 
  \int dz \dfrac{dN}{dz} \delta_g (\chi \hat{\bm n}, z),
\end{equation}
where $\delta_g$ represents the density contrast of tracers, $\hat{\bm
  n}$ is the direction to the line of sight, $dN/dz$ is a normalized
distribution function of tracers in redshift such that $\int dz dN/dz
= 1$ and $\chi (z)$ is the comoving distance to the redshift $z$.  We
assume the following analytic form of the normalized galaxy
distribution function,
\begin{equation}
  \dfrac{dN}{dz} =  
  \dfrac{\beta z^{\alpha}}{\Gamma 
    \left[(\alpha + 1)/\beta \right] z_0^{\alpha + 1}} \exp 
  \left[ -\left(\dfrac{z}{z_0} \right)^{\beta}\right] ,
\label{Gdist}
\end{equation}
where $\alpha$, $\beta$ and $z_0$ are the free parameters. In this
parameterization $\alpha$ and $\beta$ denote the slope of the
distribution at low and high-redshifts, respectively, and $z_0$
determines the peak of the distribution.  We assume that the tracer
density field is related to the underlying matter density field via a
scale- and redshift-dependent bias factor, so that $\delta_g({\bm k},
z) = b(k, z) \delta({\bm k}, z)$.  On large scales, where the mass
fluctuations are small $\delta \ll 1$, the perturbations grow
according to the linear growth rate, $\delta ({\bm k}, z) = \delta
({\bm k}) T(k) D(z)$, where $\delta({\bm k})$ is the primordial value
of matter density and $T(k)$ is the transfer function.  The linear
angular power spectrum of the galaxy distribution for a flat universe
is given by
\begin{equation}
  \cgg = \dfrac{2}{\pi} \int k^2 dk P(k) {\Delta_l^g (k)}^2 ,
\end{equation}
where
\begin{equation}
  \Delta_l^g (k) = \int dz \dfrac{dN}{dz} b(k, z) T(k) D(z) j_l ({k}\chi) .
\label{eq:delg}
\end{equation}
and $P(k)$ is the linear power spectrum as a function of the wave number $k$ 
and $j_l(k\chi)$ is a spherical Bessel function. 

In order to estimate errors in parameters and signal-to-noise
ratios, we need to describe the noise contribution due to the finiteness 
in numbers of sources associated with
source catalogs. We can write the shot noise contribution as 
\begin{equation}
  N_l^{gg} = \dfrac{1}{n_{\rm L}} ,
\end{equation}
where $n_{\rm L}$ is the surface density of sources per steradian and
related to the total number of available samples $N_{\rm g}$ as
$n_{\rm L} = N_{\rm g} / 4\pi f_{\rm sky}$.  We show the angular power
spectrum of the galaxy distribution, $\cgg$, and the noise spectrum,
$N_l^{gg}$, in the left panel of Fig.~\ref{fig:CMBNoise}.

\subsection{CMB Lensing Potential}
We consider the potential that deflects CMB photons. The relationship
between the lensed temperature anisotropy, $\tilde{T}(\hat{\bm n})$,
and unlensed one, $T(\hat{\bm n})$, is related by $\tilde{T}(\hat{\bm
  n}) = T(\hat{\bm n}+{\bm d})$ and the deflection angle ${\bm d}(
\hat{\bm n})$ is related to the line of sight projection of the
gravitational potential $\Psi (\chi \hat{\bm n}, \eta)$ as ${\bm d}(
\hat{\bm n}) = \nabla \psi ( \hat{\bm n})$, where
\begin{equation}
\psi(\hat{\bm n}) = -2 \int d\chi \dfrac{f_K(\chi_{*}) -
  f_K(\chi)}{f_K(\chi_{*}) f_K(\chi)} \Psi (\chi \hat{\bm n} ;  \eta_0 - \chi) . 
\end{equation}
Here $\psi(\hat{\bm n})$ is the lensing potential, $f_K(\chi)$ is the
angular diameter distance, $\chi$ is the radial comoving distance
along the line of sight, and $\chi_{*}$ denotes the distance to the
last scattering surface.  For a flat universe angular diameter
distance is related to the comoving distance as $f_K(\chi) = \chi$.
The angular power spectrum of the lensing potential for a flat
universe can be written as
\begin{equation}
  \cpp = \dfrac{2}{\pi} \int k^2 dk P(k) \Delta_l^{\psi} (k) ^2, 
\end{equation}
where 
\begin{equation}
  \Delta_l^{\psi} (k) = -2 \int_0^{\chi_*} d\chi T_{\Psi}(
  k ; \eta_0 - \chi) \left(\dfrac{\chi_{*} -
      \chi}{\chi_{*}\chi}\right) j_l (k\chi) . 
\label{eq:delp}
\end{equation}
In the linear theory, we define a transfer function 
for the gravitational potential $T_{\Psi}(k; \eta)$
so that $P_{\Psi}({k}; \eta) = T_{\Psi}^{2}(k; \eta) P({k})$.

The lensing potential can be reconstructed using quadratic statistics
in the temperature and polarization data that are optimized to extract
the lensing signal.  To reconstruct the lensing potential $\psi$, one
needs to use the non-Gaussian information imprinted into the
CMB. Lensing conserves surface brightness, so that the probability
distribution function of the temperatures remains unchanged. Therefore
the lowest order nonzero estimator of the lensing potential is
quadratic. This quadratic estimator has been investigated by
\cite{Hu-02, Okamoto-03} and the minimum variance estimator was given
by \cite{Hirata-03a}.  A quadratic estimator in the flat-sky
approximation generally has the form \cite{Hu-02}
\begin{equation}
  \hat{\psi}({\bm L}) = 
  N(\bm{L}) \int \dfrac{d^2{\bm l}}{(2 \pi)^2}\tilde{\Theta} ({\bm l}) 
  \tilde{\Theta}^{'} ({\bm L}-{\bm l}) {g}({\bm l}, {\bm L}),
\label{eq:lp}
\end{equation}
where $\tilde{\Theta}$ and $\tilde{\Theta}^{'}$ are lensed temperature
and/or polarization modes on the sky, $i.e.,$ $\tilde{\Theta}$,
$\tilde{\Theta}^{'} = \tilde{T}$, $\tilde{E}$, $\tilde{B}$.  The
optimal weight ${g}({\bm l}, {\bm L})$ and normalization $N(\bm{L})$
for each mode are found using the fact that the deflection position
can be written as a first order expansion of the displacement around
the undeflected position, $\tilde{\Theta} (\hat{\bm n}) = \Theta
(\hat{\bm n} + {\bm d}) = \Theta (\hat{\bm n}) + \nabla^i \psi
(\hat{\bm n}) \nabla_i \Theta (\hat{\bm n})$.  Requiring the estimator
to be unbiased and minimizing the variance, the optimal weight for
$TT$ estimator is
\begin{equation}
  {g}({\bm l}, {\bm L}) = 
  \dfrac{({\bm L} - {\bm l})\cdot{\bm L}C_{|{\bm L}-{\bm l}|}+{\bm l}
    \cdot {\bm L}C_l}{2 \tilde{C}_l^{\rm tot} 
    \tilde{C}_{|{\bm L}-{\bm l}|}^{\rm tot}} , 
\end{equation}
where $C_l$ ($\tilde{C}_l$) is the unlensed (lensed) temperature power
spectrum.  For other estimators, $C_l$ ($\tilde{C}_l$) represents the
temperature or polarization one.  The superscript "tot" originates
from the fact that the lensed CMB and the noise enter in the variance,
$\tilde{C}_l^{tot} = \tilde{C}_l + N_l$.

With the definition in Eq.~(\ref{eq:lp}), the lowest order noise 
of the lensing reconstruction equals to 
the normalization which is determined by
\begin{equation}
  \delta({\bm 0})\langle |\hat{\psi}({\bm L})|^2 \rangle = 
  N(\bm{L}) = \left[ \int \dfrac{d^2 {\bm l}}{(2\pi)^2} 
    \left[ ({\bm L} - {\bm l})\cdot{\bm L}C_{|{\bm L}-{\bm l}|}+{\bm l}
      \cdot {\bm L}C_l \right] 
    \times {g}({\bm l}, {\bm L}) \right]^{-1} .
\end{equation}
Physically the variance is a combination of the noise introduced 
by primary anisotropies themselves and the instrumental noise. 
The all-sky generalization is presented in Ref.~\cite{Okamoto-03}. 

\begin{figure}[t]
  \includegraphics[clip,keepaspectratio=true,width=0.46
  \textwidth]{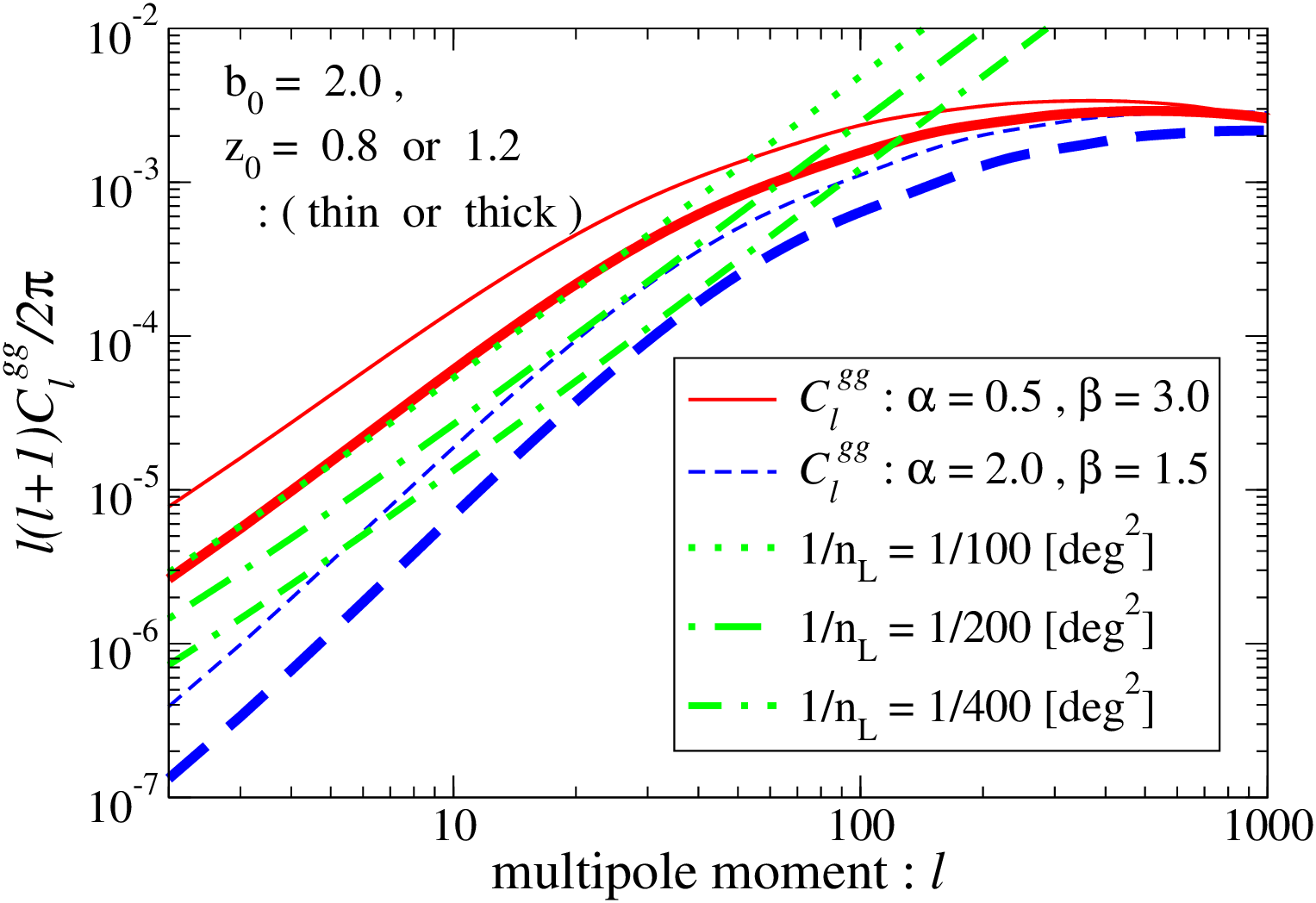}
  \hspace{5mm}
  \includegraphics[clip,keepaspectratio=true,width=0.45
  \textwidth]{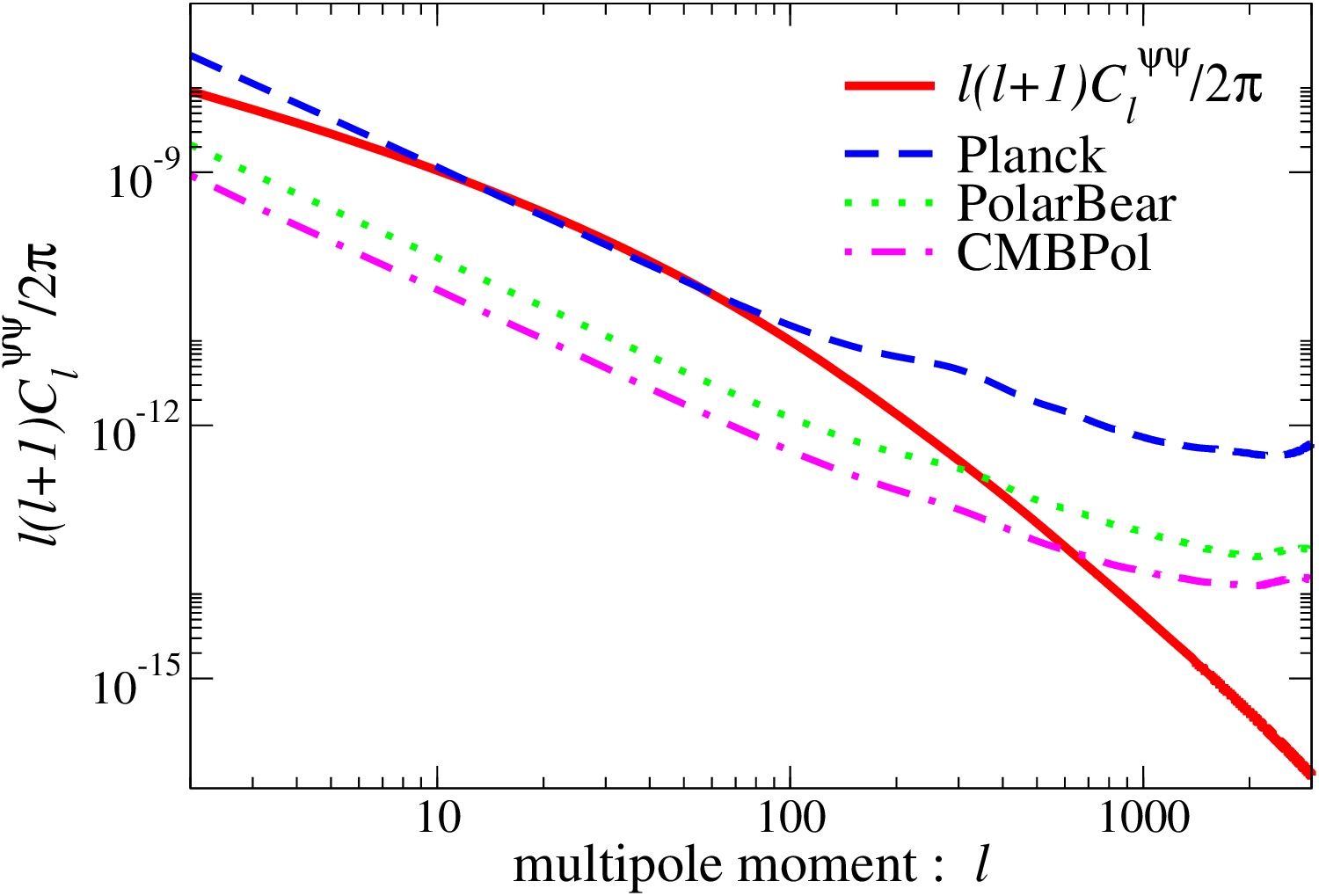}
\caption{ (Left) Angular power spectrum of the galaxy distribution,
  $\cgg$, for the Gaussian initial condition ($f_{\rm NL} = 0$), and
  the galaxy shot noise, $1/n_{\rm L}$. We show $\cgg$ for various
  forms of galaxy sampling model [Eq. )\ref{Gdist})], $(\alpha,~\beta)
  = (0.5,~3.0)$ (solid line) or ($\alpha,~\beta) = (2.0,~1.5)$ (dashed
  line), and $z_0 = 0.8$ (thin line) or $z_0 = 1.2$ (thick line).  The
  dotted, dot-dashed and dash-dot-dashed lines show shot the noise
  contributions from $n_{\rm L}$ = 100, 200, 400 $[\rm{deg}^{-2}]$,
  respectively.  (Right) Angular power spectrum of CMB lensing
  potential, $\cpp$ (solid line), and lensing reconstruction noise,
  $N_l^{\psi \psi}$ (non-solid lines).  Each of the noise power
  spectrum indicates those for Planck (dashed line), PolarBear
  (dot-dashed) and CMBPol (dot-dash-dotted line), respectively.
\label{fig:CMBNoise}}
\end{figure}

Here, the noise power spectrum of the CMB experiment reads
\begin{equation}
  N_{l,\, \nu}^{XX} = (\theta_{\rm FWHM} \Delta_X)^2 \exp 
  \left[ l(l + 1)\theta_{\rm FWHM}^2 /8 \ln 2 \right],
\end{equation}
with $X \in \{ T, E, B\}$, where $\Delta_{X}$ is the temperature and
polarization sensitivities per pixel of the combined detectors and
$\theta_{\rm FWHM}$ describes the spatial resolution of the
beam. These values are given for each frequency bands $\nu$ and we
show the values for some CMB experiments in Table
\ref{tb:CMBNoisePar}.  When there are multiple frequency bands or
$channels$, the global noise of the experiment is given by 
\begin{equation}
  N_l^{XX} = \left[ \sum_{\nu} (N_{l,\, \nu}^{XX} )^{-1} \right]^{-1},
\end{equation}
where the sum is over the individual channels.  We show the angular
pawer spectrum of the CMB lensing potential, $\cpp$, and its noise
spectrum, $N_l^{\psi \psi}$, for various CMB experiments in the right
panel of Fig.~\ref{fig:CMBNoise}.  As Planck does not have much
sensitivity to reconstruct the lensing potential from the polarization
components, $TT$ provides the best estimator for the Planck. For the
reference experiment like the CMBPol, the lensing potential, however,
is reconstructed from polarization components and $EB$ provides the
best estimator.

\begin{table}[t]
  \begin{ruledtabular}
    \begin{tabular}{lccccc} 
      Experiments & $f_{\rm sky}$  & $\nu$ [{\rm GHz}] & $\theta_{\rm FWHM}$  & $\Delta_T$  & $\Delta_P$ \\ \hline
      Planck \cite{Planck}	        & 0.65	 & 100	& 9.5$^{'}$	& 6.8 & 10.9\\ 
      & 		 & 143	& 7.1$^{'}$	& 6.0 & 11.4\\ 
      & 		 & 217	& 5.0$^{'}$	& 13.1& 26.7\\ \hline
      PolarBear  \cite{PolarBear}      & 0.03	 & 90	& 6.7$^{'}$	& 1.13 & 1.6\\ 
      & 		 & 150	& 4.0$^{'}$	& 1.70& 2.4\\ 
      & 		 & 220	& 2.7$^{'}$	& 8.00& 11.3\\ \hline
      CMBPol \cite{CMBPol}	        & 0.65	 & 100	& 4.2$^{'}$	& 0.87& 1.18\\ 
      & 		 & 150	& 2.8$^{'}$	& 1.26& 1.76\\ 
      & 		 & 220	& 1.9$^{'}$	& 1.84& 2.60\\ 
    \end{tabular}
  \end{ruledtabular}
  \caption{The current designs of CMB experiments. $\theta_{\rm FWHM}$ is 
the Gaussian beam width at FWHM, $\Delta_T$ and $\Delta_P$ are 
the temperature and polarization noises, respectively. 
Planck and CMBPol are the satellite experiments and PolarBear is 
the ground based experiment. 
\label{tb:CMBNoisePar}}
\end{table}

\subsection{Cross-Correlation: Galaxy \& Lensing Potential}
We focus on the linear cross-spectrum of the galaxy with the CMB
lensing potential, 
\begin{equation}
  C_l^{\psi g} = \dfrac{2}{\pi} \int
  k^2 dk P(k) \Delta_l^{\psi} (k) \Delta_l^g (k) .  
\end{equation}The most
important assumption we have made so far is that the galaxy
distribution and the lensing potential is linear and Gaussian.  On
small scales this will not be quite correct due to non-linear
evolution. For simple models, fits to numerical simulation like the
HALOFIT code of Ref.~\cite{Smith-03} can be used to compute an
approximate non-linear power spectrum. A good approximation is
simply to scale the transfer functions $T(k)$ of
Eq.~(\ref{eq:delg}), (\ref{eq:delp}) so that the power spectrum has
the correction from the non-linear effect 
\begin{equation}
  T(k) ~ \longrightarrow ~ T(k) \sqrt{\dfrac{P^{\rm non-linear}(k) }{P(k)}} .  
\end{equation}

\begin{figure}[t]
\centering
\includegraphics[clip,keepaspectratio=true,width=0.45
\textwidth]{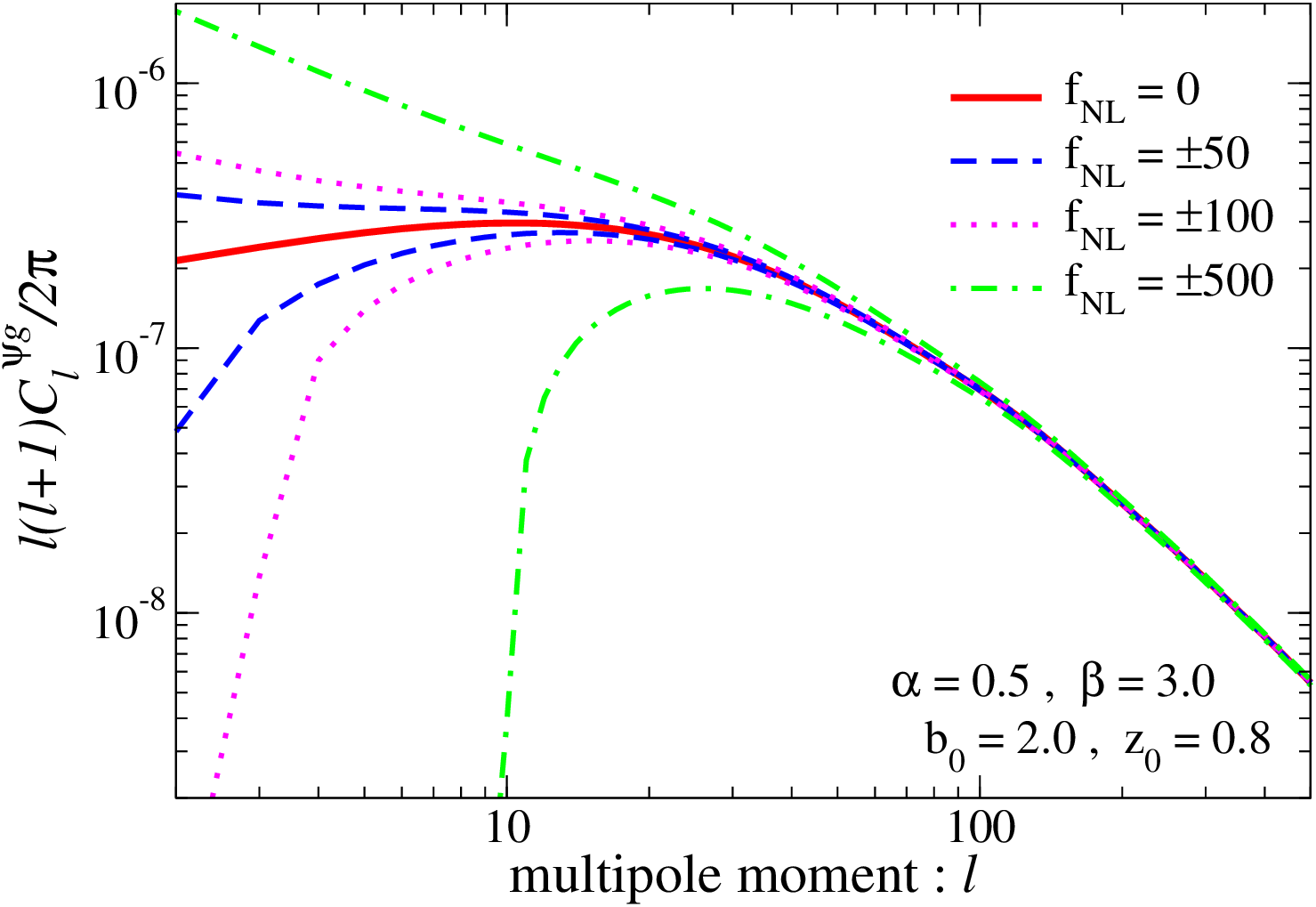}
\hspace{10mm}
\includegraphics[clip,keepaspectratio=true,width=0.45
\textwidth]{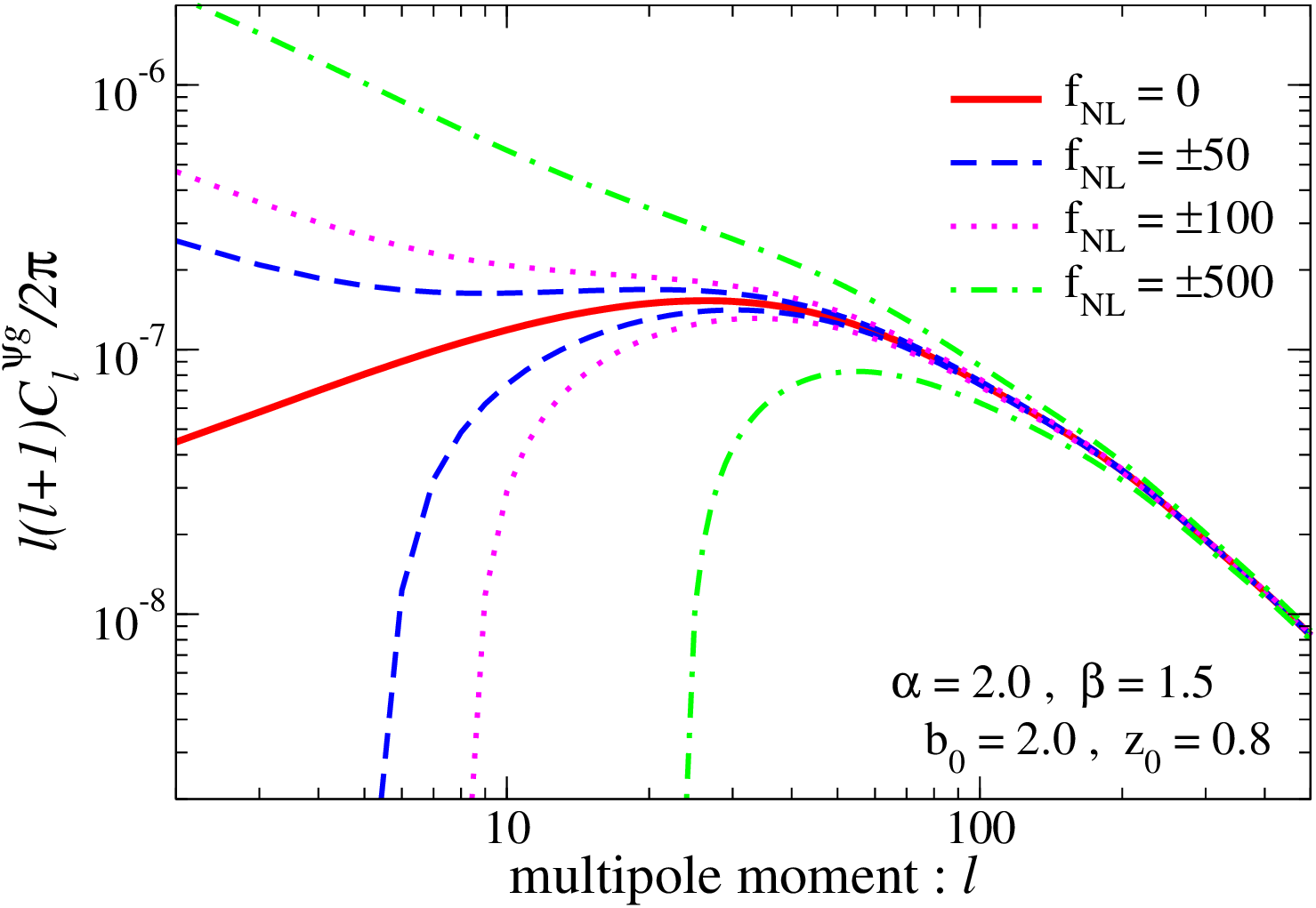}

\vspace{5mm}
\includegraphics[clip,keepaspectratio=true,width=0.45
\textwidth]{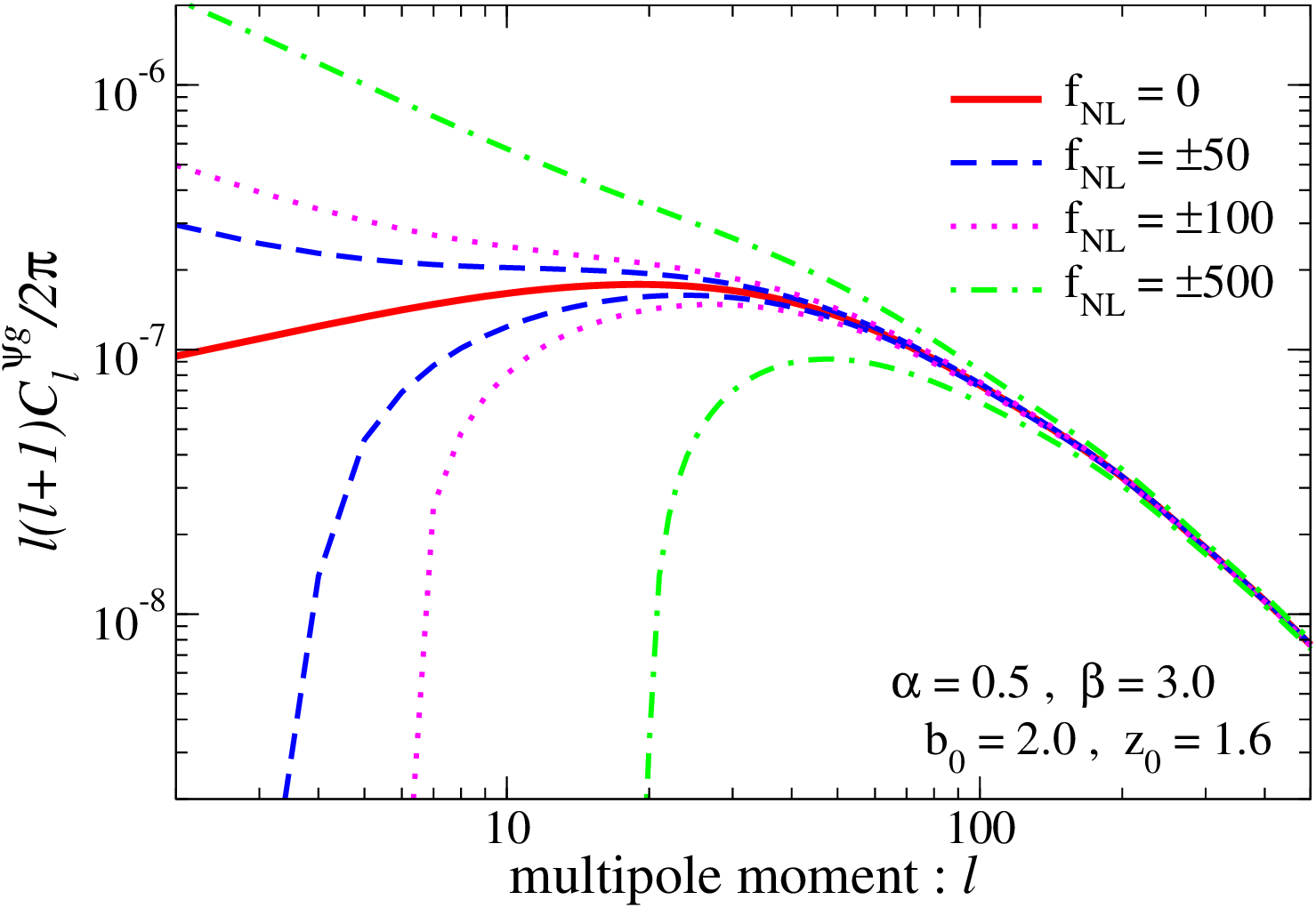}
\hspace{10mm}
\includegraphics[clip,keepaspectratio=true,width=0.45
\textwidth]{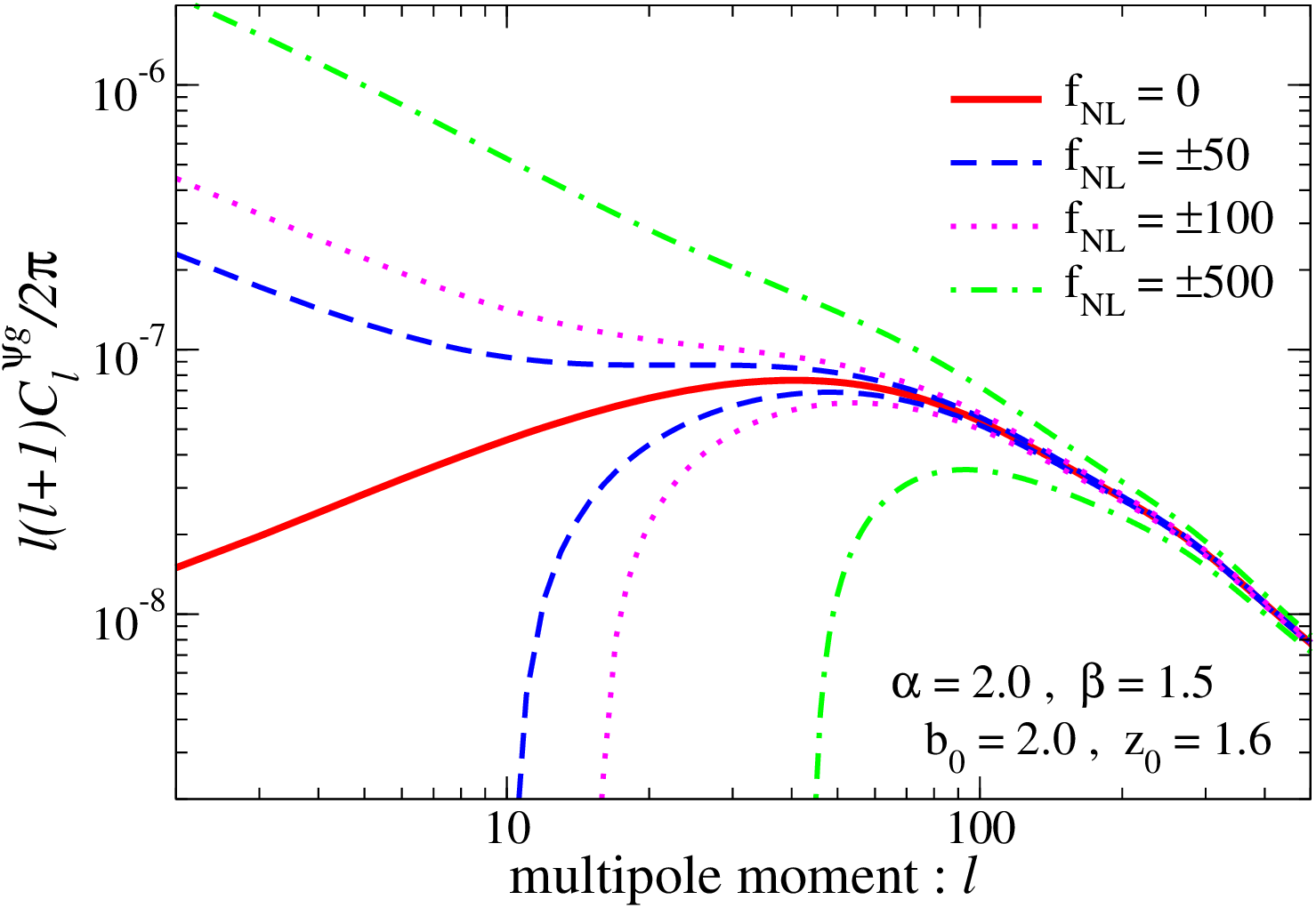}

\vspace{5mm}
\includegraphics[clip,keepaspectratio=true,width=0.45
\textwidth]{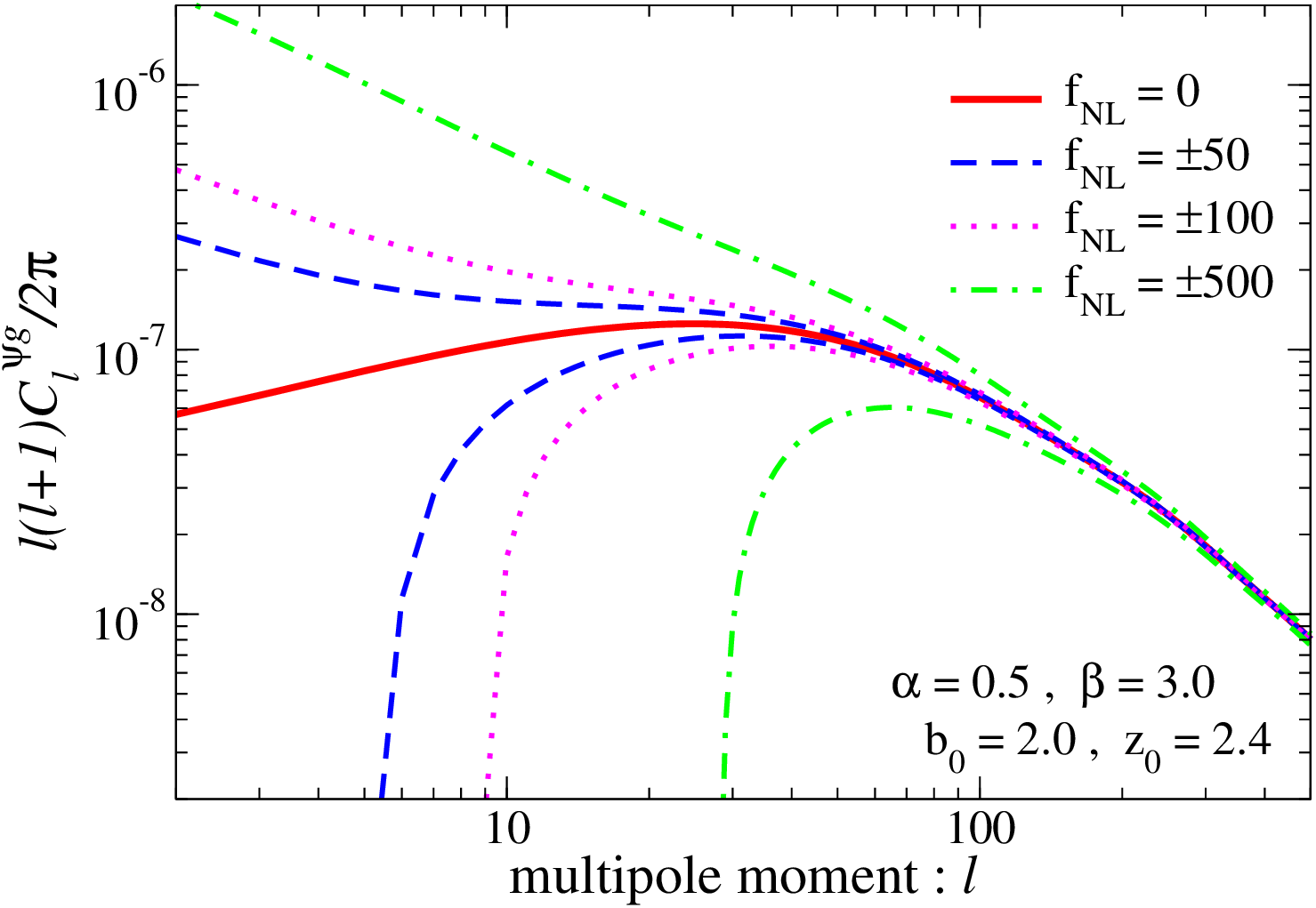}
\hspace{10mm}
\includegraphics[clip,keepaspectratio=true,width=0.45
\textwidth]{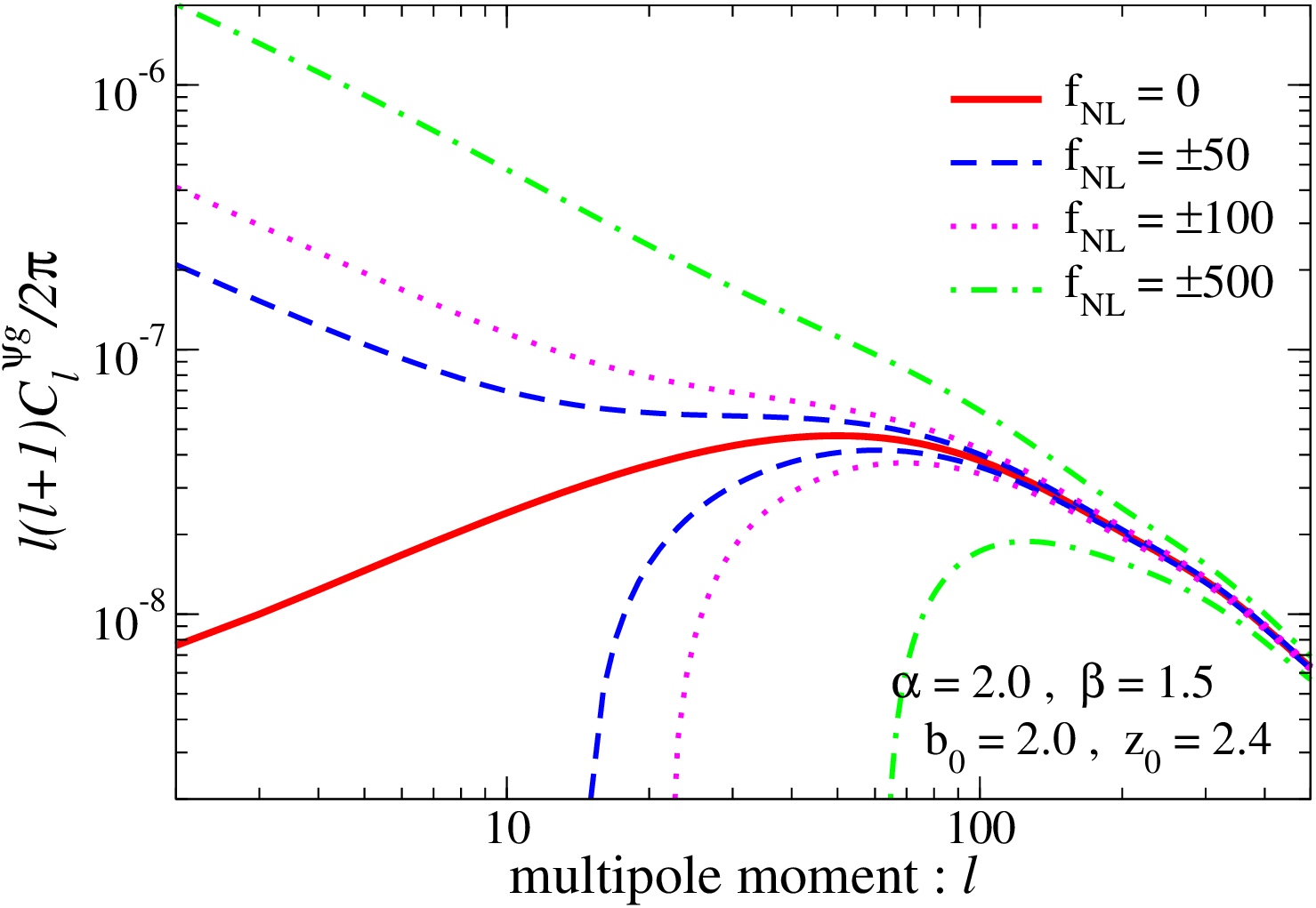}
\caption{
The cross-correlation power spectrum between the CMB lensing potential 
and the galaxy distribution, $l(l+1)C_l^{\psi g}/2\pi$ for $f_{\rm sky}=0.1$ 
and $N_{\rm g}=10^6$. The solid, dashed, dotted and dot-dashed lines correspond 
to $f_{\rm NL} = 0, \pm 50, \pm 100, \pm 500$, respectively.  
\label{fig:Clpg}}
\end{figure}

We also include the other cross-correlation components, $\cte$, $\ctp$
and $\ctg$, for the estimation of the parameter errors.  However, we
assume that there is no cross-correlation between the polarization and
the lensing potential or the galaxy distribution, $\cep = \ceg =
0$. This is because the polarization is mainly produced by the Thomson
scattering at the last scattering surface while the lensing potential
and the galaxy distribution exist in the late-time universe.  We show
the angular power spectrum $\cpg$ in Fig.~\ref{fig:Clpg}. The redshift
dependence and the effect of the primordial non-Gaussianity through a
scale-dependent bias are clearly seen in that figure.

\section{Modeling Galaxy Sample}
\label{sec:Modeling}
We showed the analytic form of the normalized galaxy distribution
function in Eq.~(\ref{Gdist}).  The mean redshift $z_{\rm m}$ is
related to the peak redshift $z_0$ and determined by
\begin{equation}
  z_{\rm m} = \int dz z \dfrac{dN}{dz} = \dfrac{z_0 \Gamma 
    \left[(\alpha + 2)/\beta \right]}{\Gamma \left[(\alpha + 1)/\beta \right]} .
\label{meanredshift}
\end{equation}
The relation between $z_0$ and $z_{\rm m}$ is, for example, $z_0 =
{z_{\rm m}} /0.64$ for ($\alpha$, $\beta$) = (0.5, 3.0) and $z_0 =
{z_{\rm m}} /1.41$ for ($\alpha$, $\beta$) = (2.0, 1.5).  In this
paper, we consider a wide field survey such as the on-going Hyper
Suprime-Cam (HSC) project. This is a fully funded imaging survey at
Subaru telescope.  The surface density $n_{\rm L}$ and the mean
redshift ${z_{\rm m}}$ are related to the exposure time $t_{\exp}$ as
\cite{Amara-06, Yamamoto-07}, 
\ba
{z_{\rm m}} &=& 0.9 \left( \dfrac{t_{\exp}}{\rm 30 min} \right)^{0.067} , \\
\label{eq:model-zm}
n_{\rm L} &=& 35 \left( \dfrac{t_{\exp}}{\rm 30 min} \right)^{0.44} \; {\rm [arcmin^{-2}]} .
\label{eq:model-nl}
\ea 
In Ref.~\cite{Amara-06} and \cite{Yamamoto-07}, ($\alpha$,
$\beta$) = (0.5, 3.0) and (2.0, 1.5) are adopted, respectively.  In
this paper, we adopt both cases and compare the differences between
the survey models.

The validity of the above form of the galaxy distribution is shown in
Ref.~\cite{Yamamoto-07}.  They compared it with the
Canada-France-Hawaii telescope (CFHT) photometric redshift data
\cite{Ilbert-06}. The relationship between magnitude limit and
exposure time was scaled for the published Subaru Suprime-Cam specification
\cite{Miyazaki-02}, and these data are shown in Table~\ref{tb:Exptime}
for the $i, g, r, z$ passbands.

The total survey area can be expressed as \cite{Yamamoto-07}
\begin{equation}
  {\rm area} = \pi \left( \dfrac{\rm field \,of\, view}{2} \right)^2 
  \dfrac{T_{\rm total}}{1.1 \times t_{\rm \exp} + t_{\rm op}} ,
\end{equation}
where we assume that the field of view is 1.5$^\circ $, the total
observation time $T_{\rm total}$ is fixed as 800 hours, and the
overhead time is modeled by constant, $t_{\rm op} = 5$ min, plus a 
fraction (10\%) of the exposure time $t_{\rm \exp}$ for one field
of view.  
\begin{table}[t]
\begin{ruledtabular}
\begin{tabular}{ccccc}
$i_{\rm AB~limit}$ & $i(S/N=10)$ & $g(S/N=5)$ & $r(S/N=5)$ & $z(S/N=5)$  \\ 
\hline 
 $22.97$ & $1$~mins.  & $3$~mins.   & $1.1$~mins. & $0.3$~mins. \\
 $23.84$ & $5$~mins.  & $15$~mins.  &  $7$~mins.  & $1.4$~mins. \\
 $24.22$ & $10$~mins. & $30$~mins.  & $12$~mins.  & $3.5$~mins. \\
 $24.81$ & $30$~mins. & $90$~mins.  & $34$~mins.  & $8.1$~mins. \\
 $25.04$ & $45$~mins. & $130$~mins. & $50$~mins.  & $13$~mins.  \\
\end{tabular}
\caption{
Exposure time for the bands, $i, g, r, z$. The relation between magnitude limit 
and exposure time was scaled for the published Subaru Suprime-Cam specification 
\cite{Miyazaki-02, Yamamoto-07}. 
\label{tb:Exptime}}
\end{ruledtabular}
\end{table}

\section{The Fisher matrix analysis} 
\label{sec:Fisher}
For our Fisher matrix analysis, we refer to the method of Ref.~\cite{Perotto-06}
and expand it to take into account the cross-correlation between the
lensing potential and the galaxy distribution. In
Ref.~\cite{Perotto-06}, the $5 \times 5$ covariance matrix is
calculated for primary CMB and CMB lensing. In our case, we expand it
into $8 \times 8$ covariance matrix for the cross-correlation between
CMB lensing and galaxy distribution.

\subsection{Likelihood Function}
Each data points have contributions from both signal and
noise.  If we assume both contributions are Gaussian distributed, we
can write the likelihood function of the data given the theoretical
model as 
\begin{equation}
{\cal L}({\bm d} |\Theta) \propto \dfrac{1}{\sqrt{{\rm det}
    \bar{C}(\Theta)}}\exp \left( -\frac{1}{2} \bm{d}^\dagger
  [\bar{C}(\Theta)^{-1}] \bm{d} \right),
\label{LikeF}
\end{equation}
where $\bm{d} = (a^T_{lm},a^E_{lm},a^{\psi}_{lm}, a^g_{lm} )$ is the
data vector, $\Theta = (\theta_1, \theta_2, \ldots)$ is a vector
describing the theoretical model parameters, and $\bar{C}(\Theta)$ is
the theoretical data covariance matrix represented by both signal and
noise. 
For it to be a good estimate, we would like it to be unbiased, $i.e.$,
$\langle \Theta \rangle = \Theta_0$ , where $\Theta_0$ indicates the
true parameter vector of the underlying cosmological model, $\Theta$
is the one constructed by the data vector $\bm{d}$ they minimizing the
likelihood function ${\cal L}({\bm d} |\Theta)$ ($i.e.$, the so-called
best fit model) and $\langle \ldots \rangle$ denotes an average over
many independent realizations.

We can derive the effective chi-square, $\chi_{\rm eff}^2
\equiv \sum_{XY} \sum_{lm}(- 2) \ln {\cal L}$, from (\ref{LikeF}) as 
\begin{equation}
\chi^2_{\rm eff} = \sum_{l} (2l+1) \left( \frac{D}{|\bar{C}|} + \ln
  |\bar{C}| \right), 
\label{effchi}
\end{equation}
where $\sum_{XY}$ represents the summation for each modes, $X, Y =
\{T, E, \phi, g\}$, and $|\bar{C}|$ denotes the determinant of the
theoretical data covariance matrix, 
\ba 
|\bar{C}| 
&=& (\bcte)^2(\bcpg)^2 -\bctt \bcee (\bcpg)^2-\bcpp \bcgg (\bcte)^2 \nn \\
&& -\bcee \bcpp (\bctg)^2 -\bcee \bcgg (\bctp)^2 +\bctt \bcee \bcpp
\bcgg +2\bcee \bctp \bctg \bcpg~. \\
\ea 
Here $D$ is defined as \ba D &=& +2 \hcte \bcte (\bcpg)^2 +2
(\bcte)^2 \bcpg \hcpg
\nonumber\\
&& -\hctt \bcee (\bcpg)^2 -\bctt \hcee (\bcpg)^2 -2 \bctt \bcee \bcpg
\hcpg
\nonumber\\
&& -\hcpp \bcgg (\bcte)^2 -\bcpp \hcgg (\bcte)^2 -2 \bcpp \bcgg \bcte
\hcte
\nonumber\\
&& -\hcee \bcpp (\bctg)^2 -\bcee \hcpp (\bctg)^2 -2 \bcee \bcpp \bctg
\hctg
\nonumber\\
&& -\hcee \bcgg (\bctp)^2 -\bcee \hcgg (\bctp)^2 -2 \bcee \bcgg \bctp
\hctp
\nonumber\\
&& +\hctt \bcee \bcpp \bcgg +\bctt \hcee \bcpp \bcgg +\bctt \bcee
\hcpp \bcgg +\bctt \bcee \bcpp \hcgg
\nonumber\\
&& +2 \hcee \bctp \bctg \bcpg +2 \bcee \hctp \bctg \bcpg +2 \bcee
\bctp \hctg \bcpg +2 \bcee \bctp \bctg \hcpg ~.  \ea In the above
expression, we have assumed that the polarization component does not
correlate with the lensing potential and galaxy distribution, so we
put $C_l^{E\psi} = C_l^{E g} = 0$.

On the other hand, the mock data covariance matrix $\hat{C}$ is given
from the simulations and defined as $\hat{C} \equiv \langle \bm{d}\,
\bm{d}^\dagger \rangle$. We can estimate the power spectrum of the
mock data through the following definition, 
\begin{equation}
  \sum_m a_{lm}^{X*}a_{lm}^{Y} = (2l+1)\hat{C}_l^{XY} ~.
\end{equation}
From Bayes' theorem, we assume (\ref{effchi}) to be the distribution
of theoretical data covariance matrix $\bar{C}(\Theta)$ when mock
covariance matrix $\hat{C}$ is given.  Then, we can account $\bar{C}$
to be a variable and $\hat{C}$ to be a constant.  All expressions
introduced so far assume a full sky coverage survey. However, real
experiments can only see a fraction of the sky. We introduce a factor
$f_{\rm sky}$, where $f_{\rm sky}$ denotes the observed fraction of
the sky in the effective $\chi^2$.  We are interested only in the
confidence levels, so the normalization factor in front of the
likelihood function (\ref{LikeF}) is irrelevant.  We normalize as
$\chi_{\rm eff}^2 = 0$ if $\bar{C} = \hat{C}$ by adding arbitrary
constant and redefine $\chi_{\rm eff}^2$ from (\ref{effchi}) as 
\begin{equation}
  \chi^2_{\rm eff} = \sum_{l} (2l+1) f_{\rm sky} \left(
    \frac{D}{|\bar{C}|} + \ln{\frac{|\bar{C}|}{|\hat{C}|}} - 4 \right) ,
\label{nchieff}
\end{equation}
and $|\hat{C}|$ denotes the determinant of 
the mock (observed) data covariance matrix, 
\ba
|\hat{C}| &=& (\hcte)^2 (\hcpg)^2
-\hctt \hcee (\hcpg)^2
-\hcpp \hcgg (\hcte)^2 \nn \\
&&
-\hcee \hcpp (\hctg)^2
-\hcee \hcgg (\hctp)^2 
+\hctt \hcee \hcpp \hcgg
+2\hcee \hctp \hctg \hcpg
~.
\ea

\subsection{Fisher Information Matrix}
\label{sec:FIM}
The Fisher matrix formalism can be used to understand how accurately
we can estimate the values of vector of parameters ${\bm \Theta}$ for a
given model from one or more data sets \cite{Tegmark-97}. 
The Fisher matrix approximates the curvature of the likelihood
function $\L$ around its maximum in a space spanned by the parameters 
$\theta$. 
The usual formula requires a slight generalization to account for the
possibility that different surveys may only partially overlap in sky
coverage as we shall show below. 
The likelihood function should peak at $\Theta \simeq \Theta_0$, and
can be Taylor expanded to second order around this value. The relevant
term at second order is the Fisher information matrix, defined as 
\be
F_{ij}\equiv \left.  - \frac{\partial^2 \ln {\mathcal L}}{\partial
    \theta_i \partial \theta_j} \right|_{\, \Theta_0}~.
\label{fisher}
\ee
From the Cramer-Rao inequality, the marginalized error 
on a given parameter $\theta_i$ is given by 
$\sigma(\theta_i) = \sqrt{(F^{-1})_{ii}}$ 
for an optimal unbiased estimator such as the maximum likelihood.

Substituting equations (\ref{LikeF}) and (\ref{nchieff}) into the
above expression, the Fisher information matrix is written by
\begin{equation}
  F_{ij} = \sum_{l=2}^{l_{\rm max}}\sum_{XX',YY'}\frac{\partial
  C_l^{XX'}}{\partial
  \theta_i}({\rm Cov}_l^{-1})_{XX'YY'}\frac{\partial C_l^{YY'}}{\partial \theta_j},
\label{cov}
\end{equation}
where $i, j$ run over the cosmological parameters, $l_{\rm max}$ is 
the maximum multipole available given the
angular resolution of the considered experiment, 
and $XX', YY'  \in \{ TT, EE, TE, \psi \psi, T\psi, gg, Tg, \psi g\}$. 
The matrix ${\rm Cov}_l$ is the power spectrum covariance matrix at the $l$-multipole,
\ba
{\rm Cov}_l=\frac{2}{(2l+1)f_{\rm sky}}\left(
  \begin{array}{cccccccc}
   \Xi_{TTTT}        & \Xi_{TTEE} & \Xi_{TTTE} & \Xi_{TT \phi \phi}        & \Xi_{TTgg}         & \Xi_{TT T\phi}        & \Xi_{TT Tg}        & \Xi_{TT \phi g} \\
   \Xi_{TTEE}        & \Xi_{EEEE} & \Xi_{TEEE} & 0                         & 0                  & 0                     & 0              & 0 \\
   \Xi_{TTTE}        & \Xi_{TEEE} & \Xi_{TETE} & 0                         & 0                  & 0                     & 0              & 0 \\
   \Xi_{TT\psi \psi} & 0          & 0          & \Xi_{\psi \psi \psi \psi} & \Xi_{\psi \psi gg} & \Xi_{T\psi \psi \psi} & \Xi_{Tg \psi \psi} & \Xi_{\psi \psi \psi g}\\
   \Xi_{TTgg}        & 0          & 0          & \Xi_{\psi \psi gg}        & \Xi_{gggg}         & \Xi_{T\psi gg}        & \Xi_{Tggg}         & \Xi_{\psi g gg}\\
   \Xi_{TTT\psi}     & 0          & 0          & \Xi_{T\psi \psi \psi}     & \Xi_{T\psi gg}     & \Xi_{T\psi T\psi}     & \Xi_{T\psi Tg}       & \Xi_{T\psi \psi g}     \\
   \Xi_{TTTg}        & 0          & 0          & \Xi_{Tg \psi \psi}        & \Xi_{Tg Tg}        & \Xi_{T\psi Tg}        & \Xi_{Tg \psi g}    & \Xi_{Tg \psi g}      \\
   \Xi_{TT\psi g}    & 0          & 0          & \Xi_{\psi \psi \psi g}    & \Xi_{\psi g gg}    & \Xi_{T\psi \psi g}    & \Xi_{Tg \psi g}    & \Xi_{\psi g \psi g}     
  \end{array}
 \right)~,
 \label{eq:Cov}
 \ea 
where the auto correlation coefficients are given by
\begin{subequations}
\begin{eqnarray}
\Xi_{TTTT} 
	&=& (\ctt)^2 + \dfrac{2 (\cte)^2 \left[ \cpp (\ctg)^2 + \cgg (\ctp)^2 -2 \ctp \ctg \cpg \right]}{\left[ (\cpg)^2 - \cpp \cgg \right]} , \\
\Xi_{EEEE} 
	&=& (\cee)^2 , \\
\Xi_{TETE} 
	&=& \dfrac{1}{2}\left[ (\cte)^2 + \ctt \cee \right] + \dfrac{\cee \left[ \cpp (\ctg)^2 + \cgg (\ctp)^2 -2 \ctp \ctg \cpg \right]}{2 \left[ (\cpg)^2 - \cpp \cgg \right]} , \\
\Xi_{\psi \psi \psi \psi} 
	&=& (\cpp)^2  , \\  \nn \\
\Xi_{gggg} 
	&=& (\cgg)^2 , \\
\Xi_{T\psi T\psi} 
	&=& \dfrac{1}{2}\left[ (\ctp)^2 + \ctt \cpp \right] - \dfrac{\cpp (\cte)^2}{2 \cee} , \\
\Xi_{TgTg} 
	&=& \dfrac{1}{2}\left[ (\ctg)^2 + \ctt \cgg \right] - \dfrac{\cgg (\cte)^2}{2 \cee} , \\
\Xi_{\psi g \psi g} 
	&=& \dfrac{1}{2}\left[ (\cpg)^2 + \cpp \cgg \right] ,
\end{eqnarray}
\end{subequations}
while the cross-correlation ones are
\begin{subequations}
\begin{eqnarray}
\Xi_{TTEE} 
	&=& (\cte)^2 , \\
\Xi_{TTTE} 
	&=& \ctt \cte + \dfrac{\cte \left[ \cpp (\ctg)^2 + \cgg (\ctp)^2 -2 \ctp \ctg \cpg \right]}{\left[ (\cpg)^2 - \cpp \cgg \right]} , \\
\Xi_{TT\psi \psi} 
	&=& (\ctp)^2 , \\
\Xi_{TTT\psi} 
	&=& \ctp \left[ \ctt - \dfrac{(\cte)^2}{\cee} \right] , \\
\Xi_{TTgg} 
	&=& (\ctg)^2 , \\
\Xi_{TTTg} 
	&=& \ctg \left[ \ctt - \dfrac{(\cte)^2}{\cee} \right] , \\
\Xi_{TT\psi g} 
	&=& \ctp \ctg , \\ \nn \\
\Xi_{TEEE} 
	&=& \cte \cee , \\ \nn \\
\Xi_{T\psi \psi \psi} 
	&=& \ctp \cpp , \\ \nn \\
\Xi_{T\psi gg} 
	&=& \ctg \cpg , \\
\Xi_{T\psi Tg} 
	&=& \dfrac{1}{2}\left( \ctp \ctg + \ctt \cpg \right) - \dfrac{(\cte)^2 \cpg}{2 \cee} , \\
\Xi_{T\psi \psi g} 
	&=& \dfrac{1}{2}\left( \ctp \cpg + \ctg \cpp \right) , \\ \nn \\
\Xi_{Tg \psi \psi} 
	&=& \ctp \cpg , \\ \nn \\
\Xi_{Tg gg} 
	&=& \ctg \cgg , \\ \nn \\
\Xi_{Tg \psi g} 
	&=& \dfrac{1}{2}\left( \ctg \cpg + \cgg \ctp \right) , \\ \nn \\
\Xi_{\psi \psi gg} 
	&=& (\cpg)^2 , \\ \nn \\
\Xi_{\psi \psi \psi g} 
	&=& \cpp \cpg , \\ \nn \\
\Xi_{\psi g gg} 
	&=& \cgg \cpg .
\end{eqnarray}
\end{subequations}

\section{Forecasts}
\label{sec:Forecasts}
We estimate the parameter errors for Planck satellite using Fisher
analysis following the method introduced in Sec.
\ref{sec:Fisher}. 
Our fiducial cosmology is based on the WMAP 7-year result
\cite{Komatsu-10} within a flat, $\Lambda$CDM model framework.

The fiducial model parameters we consider are given as : 
\ba
\Theta &=& 
\{ ~\Omega_b h^2, ~\Omega_c h^2, ~\Omega_{\Lambda}, ~\tau, ~f_\nu, ~Y_{\rm He}, 
~n_s, ~\Delta_{\rm R}^{2}(k_0) \times 10^{9}, ~w_0, ~N_{\rm eff}, ~\alpha_s, 
~f_{\rm NL}, ~b_0 ~\} \nn \\
&=& \{ 0.02258, ~0.1109, ~0.734, ~0.088, ~0.02, ~0.24, ~0.963, ~2.45,
-1.0, ~3.0, ~0, ~0, ~2.0 \} , 
\ea 
where $\Omega_b$, $\Omega_c$ and
$\Omega_{\Lambda}$ are the density parameters for baryon, cold dark
matter and dark energy, respectively, $h$ is the Hubble constant,
$\tau$ is the Thomson scattering optical depth to the last scattering
surface, $f_\nu$ is the mass density of the massive neutrino relative
to the total matter density:
$f_\nu \equiv \Omega_\nu /\Omega_m$, $Y_{\rm He}$ is the primordial
helium fraction, $n_s$ is spectral index of the primordial power
spectrum, $\Delta_{\rm R}^{2}(k_0)$ is the amplitude of the primordial
power spectrum normalized at $k_0 = 0.002 ~{\rm Mpc^{-1}}$, $w$ is the
equation of state parameter of dark energy, $N_{\rm eff}$ is the
effective number of neutrinos, $\alpha_s$ is the running index,
$f_{\rm NL}$ is the non-linear parameter which represents the
primordial non-Gaussianity and $b_0$ is the linear bias parameter.
Because we assume a flat universe the Hubble parameter is adjusted to
keep our universe flat when we vary $\Omega_\Lambda$.  For neutrino
parameters, we assume the standard three neutrino species. In our
analysis, the non-linear parameter $f_{\rm NL}$ and the linear bias
parameter $b_0$ are determined by the galaxy surveys only, and the CMB
experiment plays a role in breaking the parameter degeneracies.  We
use CAMB code \cite{Lewis-00} and HALOFIT code \cite{Smith-03} to
calculate the angular power spectrum $C_l^{XY}$ and the non-linear
region of the angular power spectrum of the galaxy distribution and
lensing potential.

\begin{table}[t]
\begin{ruledtabular}
\begin{tabular}{@{\hspace{5mm}}c@{\hspace{5mm}}|p{2pt}ccccp{2pt}cp{2pt}ccc@{\hspace{5mm}}}
 &&
\multicolumn{4}{c}{CMB} && 
 galaxy &&
\multicolumn{3}{c}{CMB $\times$ galaxy}  \\ 
\cline{3-6} \cline{8-8} \cline{10-12} \vspace{-2mm}\\ 
$C_l^{XY}$ && $\ctt$ & $\cee$ & $\cte$ & $\cpp$ && $\cgg$ && $\ctp$ & $\ctg$ & $\cpg$  \\
\hline \vspace{-2mm} \\
$l_{\rm max}$ && 2500 & 2500 & 2500 & 1000 && 1000 && 1000 & 1000 & 1000 \\
$f_{\rm sky}$ && 0.65 & 0.65 & 0.65 & 0.65 && 0.10 && 0.10 & 0.10 & 0.10 \\
\end{tabular}
\caption{Survey parameters for our calculation. 
\label{tb:survey} }
\end{ruledtabular}
\end{table}

In our estimation, we include the information from temperature
anisotropies, E-mode polarization and reconstructed lensing potential. 
The range of multipoles are $2 \leq l \leq 2500$ for $\ctt$ and
$\cee$ and $2 \leq l \leq 1000$ for $\cpp$, respectively, and survey
area is taken as $f_{\rm sky}^{\rm CMB} = 0.65$ for CMB survey.  On
the other hand, we include the information from galaxy survey, $\cgg$,
where the range of multipoles are $2 \leq l \leq 1000$ and survey area
is $f_{\rm sky}^{\rm galaxy} = 0.10$. We assume that there is no
correlation between different patches, so that the area where there is a 
correlation between CMB and galaxy survey corresponds to the galaxy
survey area $f_{\rm galaxy}$. We summarize the values
we used mainly in  the following calculation in Table~\ref{tb:survey}.

The non-linear effect on the angular spectrum of the galaxy
distribution begins to appear at $l \geq 100$. As the calculation of
Fisher matrix assumes that all fields are random, the non-linear region 
of the galaxy distribution is
inadequate for this calculation.  However, because the auto
correlation signal of the galaxy distribution is dominated by the
noise term in this region as seen in Fig.~\ref{fig:CMBNoise} ($Left$),
little information of the galaxy distribution from this region can be 
expected. Therefore, we neglect the non-linear
effect of the angular power spectrum of the galaxy distribution in
this paper.

Usually, the full Fisher matrix for joint experiment of galaxy survey
and CMB is obtained simply by adding each Fisher matrices: $F_{ij} =
F_{ij}^{CMB} + F_{ij}^{\rm galaxy}$. This method, however, does not
include all the available information because it does not account for
the cross-correlation of temperature-galaxy $\ctg$ and lensing
potential-galaxy $\cpg$. To use the angular power spectrum $\ctg$ and
$\cpg$ for the estimation, and make the most of the available
information, we consider all of the conceivable cross-correlations and
calculate the full covariance matrix, which in this case is $8 {\rm
  \times} 8$ matrix while we assume that $\cep$ and $\ceg$ are not
correlated. Here, in order to account for the difference of the each
survey area, and to investigate the significance of the
cross-correlation signal, we calculate the full Fisher matrix of the
following form,

\begin{eqnarray} 
{\rm Case~(i)}~:~F_{ij}
  &=& \sum_{l = 2}^{1000}F_{ij} \{
  f_{\rm sky}^{\rm CMB}:\ctt, \cee, \cte, \cpp, \ctp \} \nn
  + \sum_{l = 2}^{1000}F_{ij} \{ f_{\rm sky}^{\rm galaxy} :\cgg  \} \\
  &+& \sum_{l = 1001}^{2500}F_{ij} \{ f_{\rm sky}^{\rm CMB}:\ctt, \cee, \cte \} , \label{eq:case1} \\
  {\rm Case~(ii)}~:~F_{ij}
  &=& \sum_{l = 2}^{2500}F_{ij} \{ f_{\rm
    sky}^{\rm CMB}:\ctt, \cee, \cte \} \nn
  + \sum_{l = 2}^{1000}F_{ij} \{ f_{\rm sky}^{\rm galaxy} :\cpp, \cgg, \cpg  \} \label{eq:case2} \\
  &+& \sum_{l = 2}^{1000}F_{ij} \{ f_{\rm sky}^{\rm CMB} - f_{\rm sky}^{\rm galaxy} :\cpp \} , \\
  {\rm Case~(iii)}~:~F_{ij}
  &=& \sum_{l = 2}^{1000}F_{ij} \{ f_{\rm sky}^{\rm galaxy}:\ctt, \cee, \cte, \cpp, \ctp, \cgg, \ctg, \cpg \} \nn \\
  &+& \sum_{l = 2}^{1000}F_{ij} \{f_{\rm sky}^{\rm CMB} - f_{\rm
    sky}^{\rm galaxy}:\ctt, \cee, \cte \} +
  \sum_{l = 1001}^{2500}F_{ij} \{ f_{\rm sky}^{\rm CMB}:\ctt, \cee,
  \cte \} , \label{eq:case3} 
\end{eqnarray} 
where "$f_{\rm sky}^{(\rm Survey)}$ " represents the available sky
fraction in the Fisher matrix with $({\rm Survey}) = \{{\rm galaxy},
{\rm CMB}\}$, and "$C_l^{XY} , \cdot \cdot \cdot$ " denotes the
angular power spectra included in the Fisher matrix with $X, Y = \{ T,
~E, ~\psi, ~g \}$.  Case (i) does not include the cross-correlations
between CMB and galaxy survey, $\ctg$ and $\cpg$, and Case (ii) does
not include the cross-correlations between temperature and galaxy or
lensing potential, $\ctg$ and $\ctp$, while Case (iii) takes account
of all cross-correlations.  We shall compare the difference in these
three cases in the following section.

\section{Result}
\label{sec:Result}

\begin{figure}[t]
\centering
\includegraphics[clip,keepaspectratio=true,width=0.4
\textwidth]{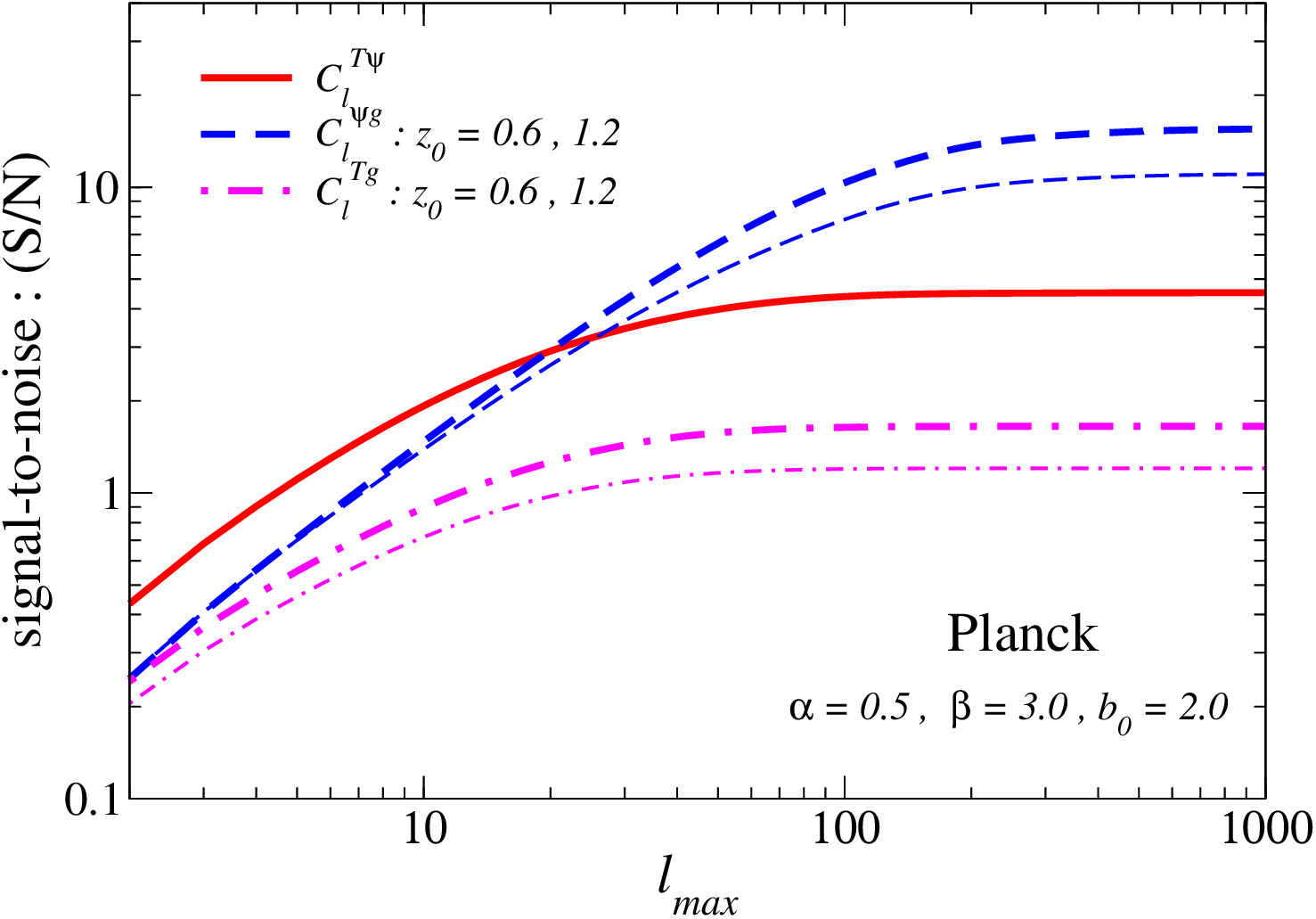}
\hspace{5mm}
\includegraphics[clip,keepaspectratio=true,width=0.4
\textwidth]{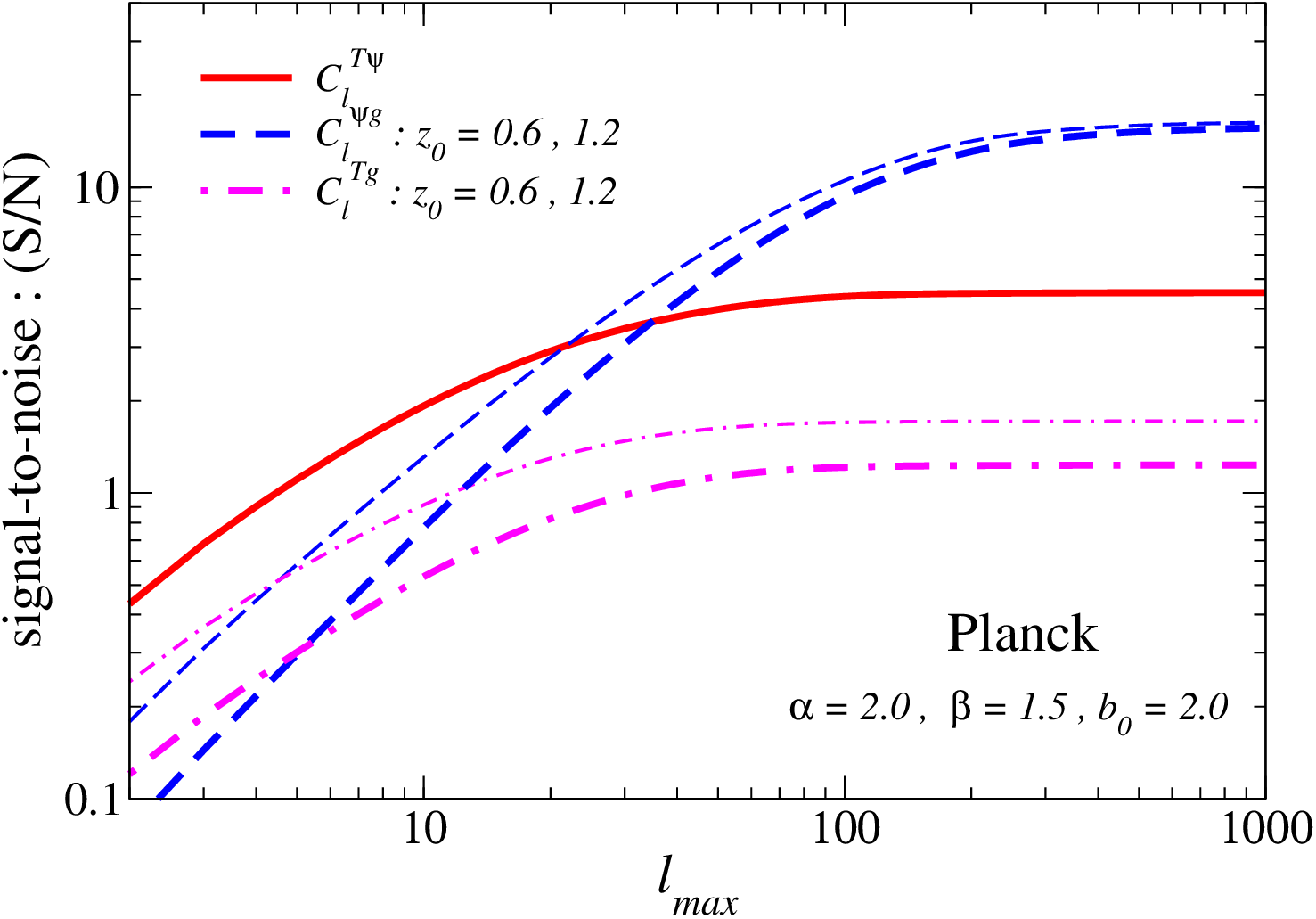}

\vspace{5mm}
\includegraphics[clip,keepaspectratio=true,width=0.4
\textwidth]{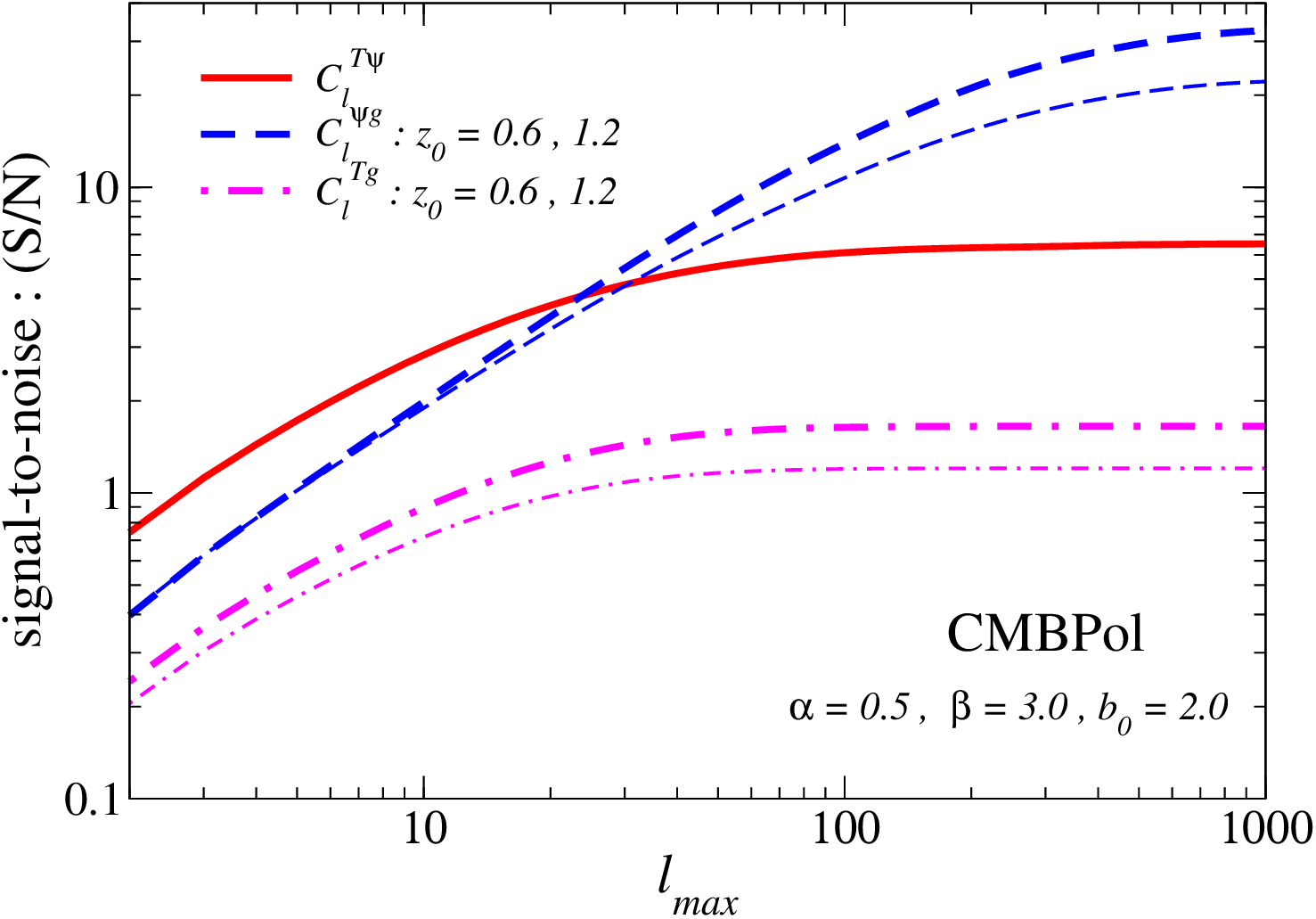}
\hspace{5mm}
\includegraphics[clip,keepaspectratio=true,width=0.4
\textwidth]{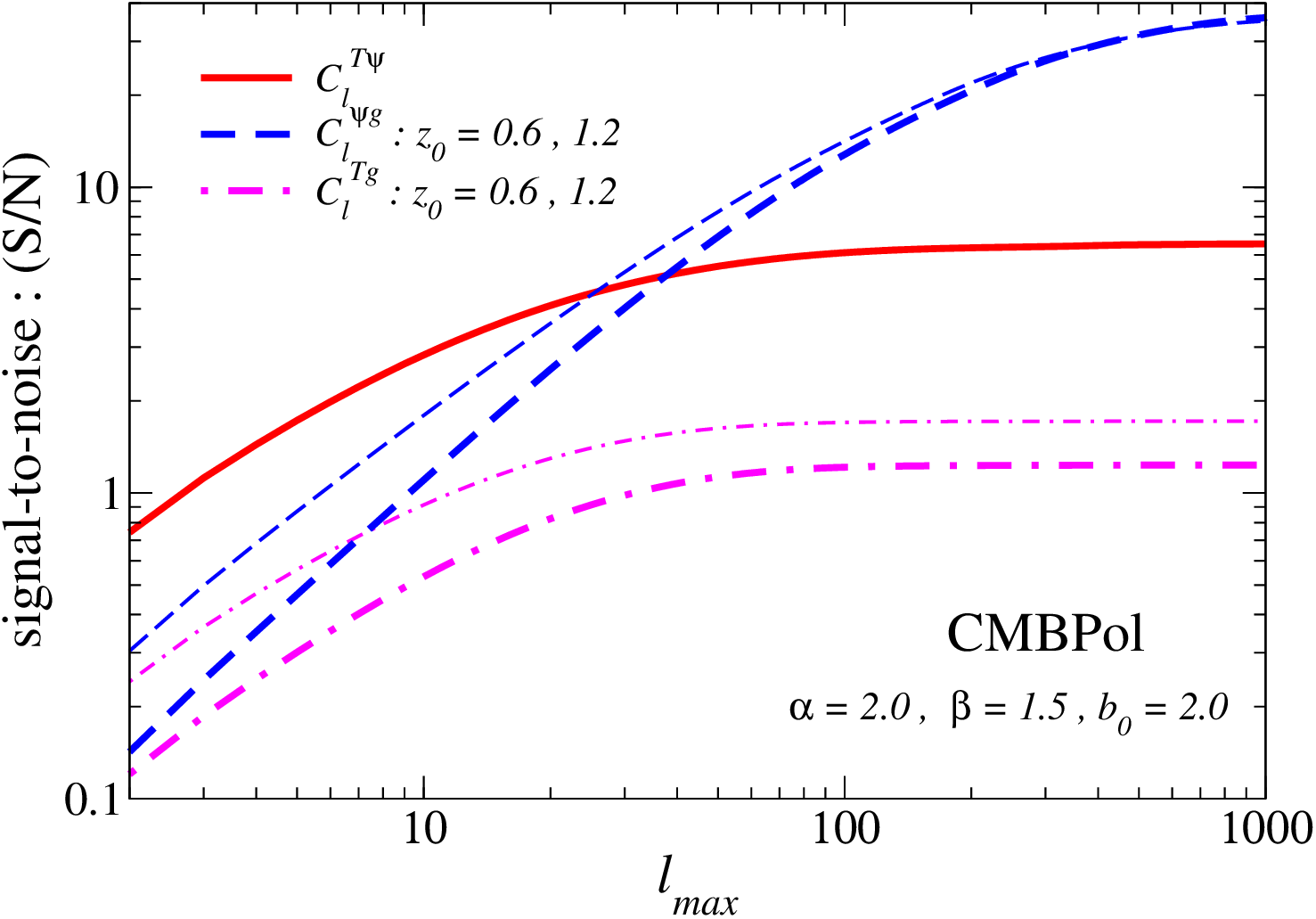}
\caption{The signal-to-noise from $\ctp$, $\ctg$ and $\cpg$, for $z_0$
  = 0.8 (thin) and 1.2 (thick), respectively, with Planck (top) and
  CMBPol (bottom). The survey area and total number of the galaxies
  are fixed to $f_{\rm sky} = 0.10$ and $N_{\rm g} = 10^6$, respectively.
  \label{fig:SN}}
\end{figure}

\subsection{Signal-to-Noise}
The signal-to-noise ratio ($S/N$) for a cross-correlation between $X$
and $Y$ can be estimated by \cite{Peiris-00} 
\begin{equation}
  \left( \dfrac{S}{N}
  \right)^2 = f_{\rm sky} \sum_{2}^{l_{\rm max}} (2 l+1)
  \dfrac{(C_l^{XY})^2}{(C_l^{XY})^2 + (C_l^{XX} + N_l^{XX})(C_l^{YY} +
    N_l^{YY})} .  
\end{equation}
We show the $S/N$ value in Fig.~\ref{fig:SN} where
we fixed the survey area as $f_{\rm sky} = 0.10$ and total number of
the galaxies as $N_{\rm g} = 10^6$, respectively. The
cross-correlations between temperature and lensing potential $\ctp$
and temperature and galaxy $\ctg$ are through the Integrated
Sachs-Wolfe (ISW) effect imprinted in the CMB and the distribution of
matter at late time. The ISW effect arise from the time-variation of
the scalar metric perturbations and it is usually divided into an
early ISW effect and a late ISW effect.  The origin of the late ISW
effect is from the time variation of the gravitational potential by
the dark energy component and its effect emerges at low-multipoles.
Therefore, the $S/N$ both from $\ctp$ and $\ctg$ saturate around
$l_{\rm max} \simeq 40$, although their amplitudes are different.

On the other hand, the cross-correlation between lensing potential and
galaxy has another feature. The survey which can explore the small
scale region with large $l_{\rm max}$, can get large $S/N$ from $\cpg$
more than those from $\ctp$ and $\ctg$ while their amplitude is very
small for the low resolution survey with small $l_{\rm max}$.  The
saturation of $S/N$ from $\cpg$ at large $l_{\rm max}$ for Planck in
Fig.~\ref{fig:SN} (top two panels) is due to the noise contribution of
the surveys.  This justifies our omitting the proper modeling on small
scales where non-linear evolutions are important.  The signal-to-noise
may be improved for high-precision future CMB survey, such as CMBPol,
because it will be obtain much more information about small scale
region than Planck.  Since the cross-correlation between lensing
potential and galaxy $\cpg$ has larger $S/N$ than other
cross-correlation components, $\cpg$ would be more powerful tool when
small-scale powers are observed.  However, it should be noted that the
correct non-linear model will be necessary on these scales.

\subsection{Parameter errors}
\begin{figure}[t]
\centering
\includegraphics[clip,keepaspectratio=true,width=1.0\textwidth]{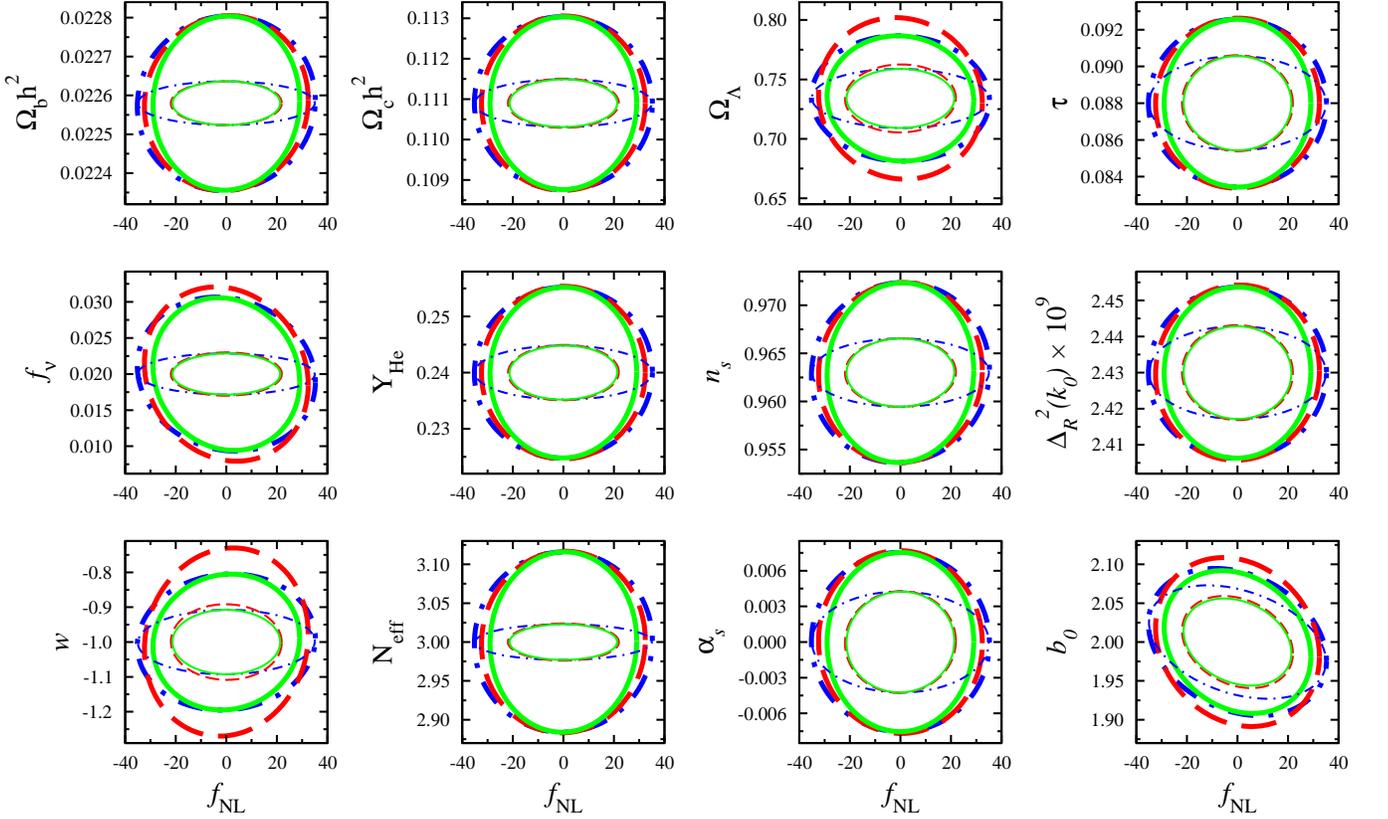}
\caption{ 1$\sigma$ confidence limits on the pair ($f_{\rm NL}$,
  $\theta_i$) in our 13-dimensional model. We show for Case (i)
  (dot-dashed line), Case (ii) (dashed line) and Case (iii) (solid
  line), respectively. We assume Planck (thick line) and CMBPol (thin
  line) for CMB survey, and for the galaxy survey the sky coverage and
  total number of galaxy sampling of $f_{\rm sky} = 0.1$ and $N_{\rm
    g} = 10^6$.  The parameters of galaxy sampling model are fixed as
  $\alpha = 0.5$, $\beta = 3.0$ and $z_0 = 1.8$.
\label{fig:ErrorCont}}
\end{figure}

We show the 1$\sigma$ marginalized error of each parameter in
Table~\ref{tb:ParError} and error contours in
Fig.~\ref{fig:ErrorCont}.  First, we compare the analysis methods,
$i.e.$, the Cases (i) - (iii) defined in Eqs.~(\ref{eq:case1}) -
(\ref{eq:case3}).  From the figure we find that the cosmological
parameters, such as $\Omega_{\Lambda}$, $f_{\nu}$, and $w$ are tightly
constrained in Case (i), while the constraint on the non-Gaussianity
parameter $f_{\rm NL}$ is tighter in Case (ii) than Case (i). Because
the main difference between Case (i) and (ii) is whether one includes
$\ctp$ or $\cpg$, respectively, we can conclude that the
cross-correlation between lensing potential and galaxy $\cpg$ is
important to determine $f_{\rm NL}$ more precisely. Assuming more
high-precision CMB survey, CMBPol, the significance of $\cpg$ for
constraints on $f_{\rm NL}$ are seen more clearly.

\begin{figure}[th]
\centering
\includegraphics[keepaspectratio=true,width=0.45
\textwidth]{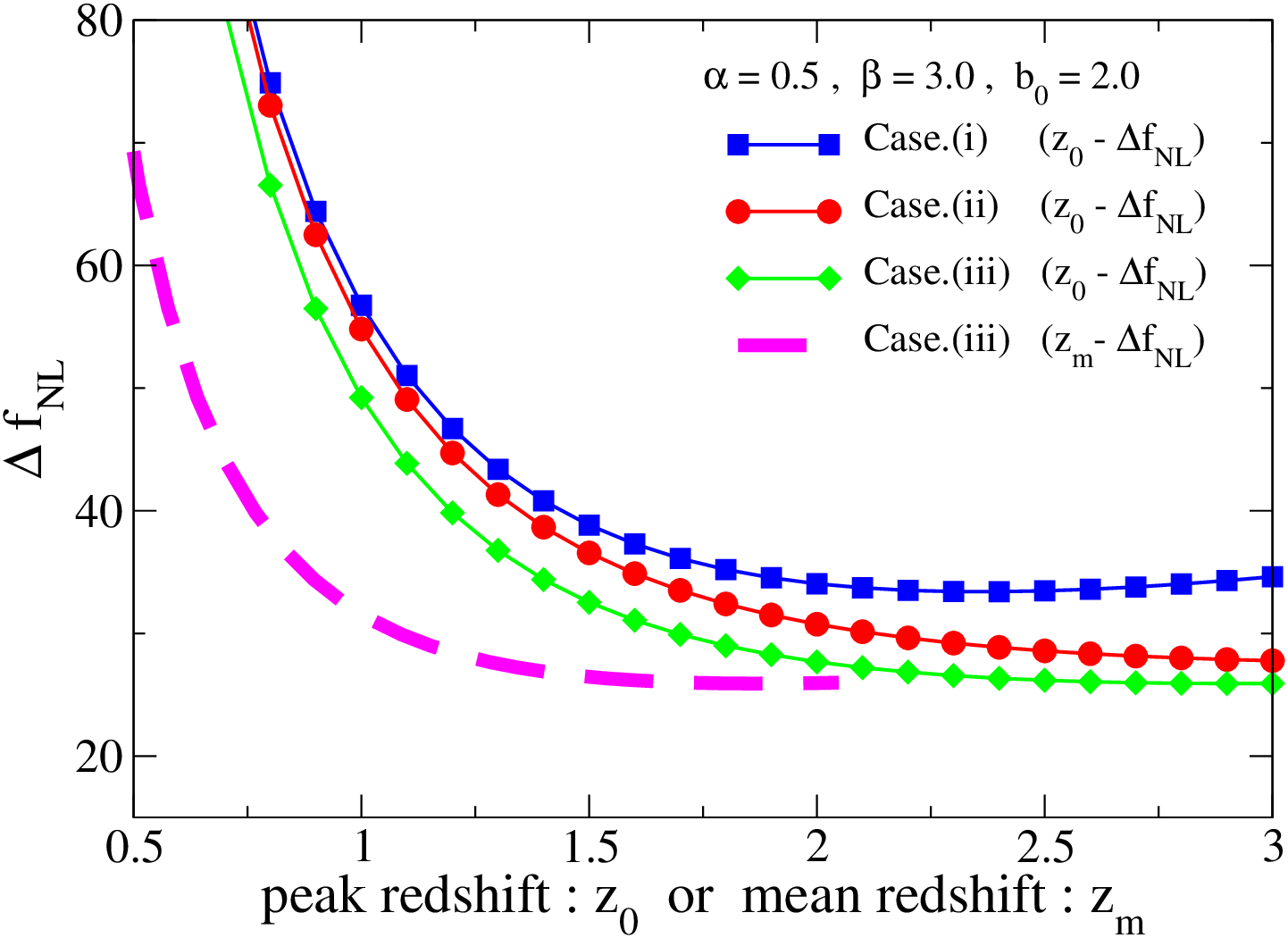}
\hspace{5mm}
\includegraphics[keepaspectratio=true,width=0.45
\textwidth]{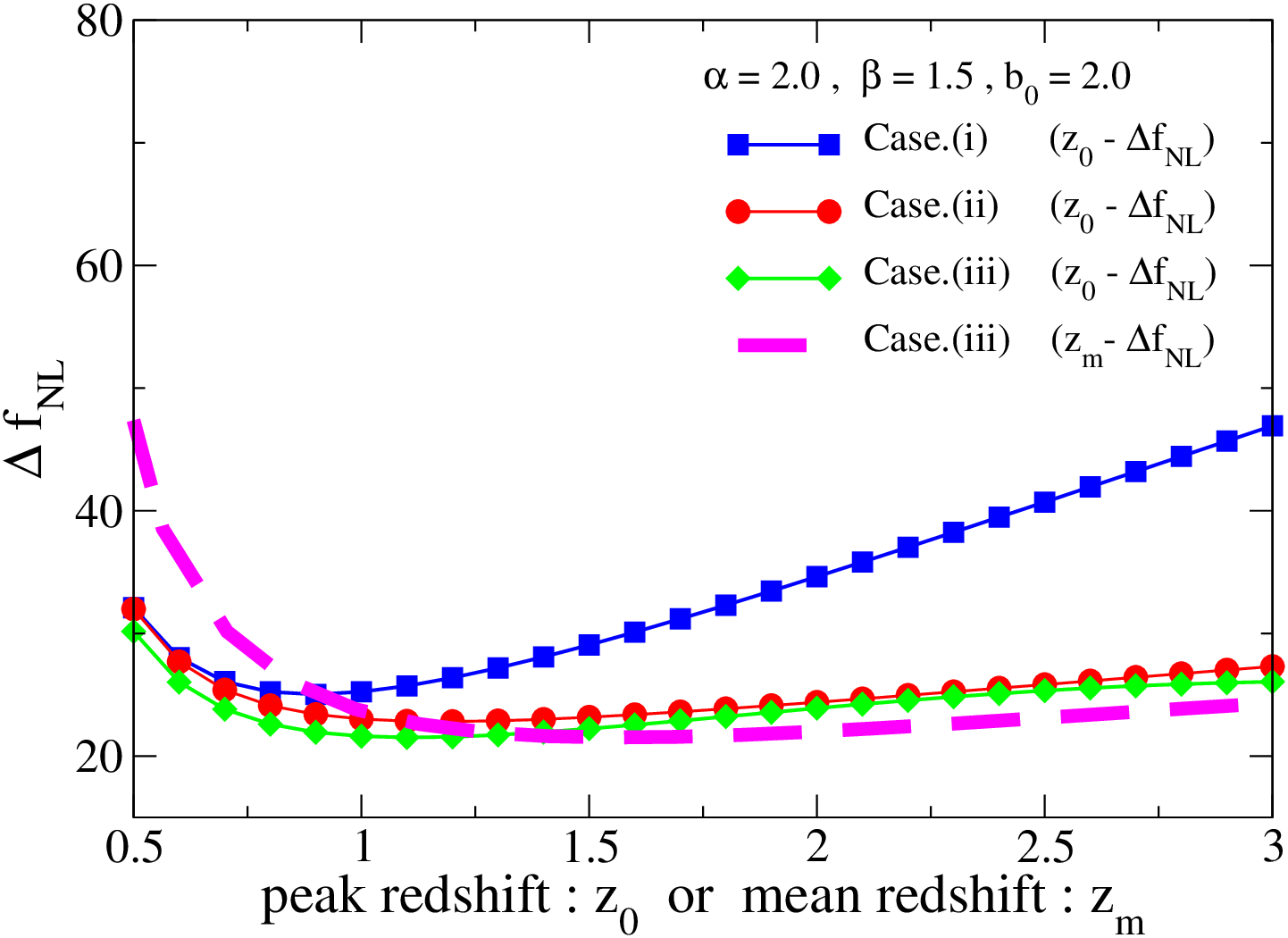}
\vspace{5mm}
\caption{ The comparison of the 1$\sigma$ error of $f_{\rm NL}$ in the
  cases with and without cross-correlations. The lines with simbols
  are against peak redshift $z_0$ and thick dashed line corresponds to
  the Case (iii) against mean redshift ${z_{\rm m}}$, respectively.
  The left panel is for the case $\alpha = 0.5$, $\beta = 3.0$ and the
  right one is for the case $\alpha = 2.0$, $\beta = 1.5$, in
  Eq.(\ref{Gdist}). Planck is considered for the CMB experiment and
  the galaxy survey parameters are taken as $f_{\rm sky}$ = 0.10 and
  $N_{\rm g} = 10^6$.
\label{fig:ParErrorComp}}
\end{figure}

Next, we investigate how the different galaxy sampling models affect
the determination of $f_{\rm NL}$.  For this purpose we consider two
cases, the cases ($\alpha$, $\beta$) = (0.5, 3.0) and ($\alpha$,
$\beta$) = (2.0, 1.5), and the results are shown in
Fig.~\ref{fig:ParErrorComp}.  The features of the two models are that
the former has a gradual distribution and the latter has a sharp one
around the peak redshift $z_0$.  Generally, the expected errors
rapidly decrease with redshift $z_0$ for $z_0 \lesssim 1$. As for the
Case (iii) the error is almost independent of $z_0$ for the case with
($\alpha$, $\beta$) = (2.0, 1.5), while it has strong dependence for
the case with ($\alpha$, $\beta$) = (0.5, 3.0).  However, translating
the peak redshift $z_0$ to the mean redshift ${z_{\rm m}}$, the
tendency of the two models is similar to each other, although the
constraint from the sharp distribution model is somewhat stronger than
the gradual one, where the mean redshift ${z_{\rm m}}$ is determined
by Eq.~(\ref{meanredshift}) and the relation between ${z_{\rm m}}$ and
$z_0$ depends on the galaxy sampling model. For the same value of the
peak redshift $z_0$, the gradual model represents relatively low mean
redshift observation and the sharp model represents high redshift
one. Therefore, the mean redshift of galaxies, rather than their
distribution, determines how tight constraint one can obtain.

The difference of the constraints due to sampling models can be
attributed to the degree of the correlation between the lensing
potential and the galaxy distribution.  Because the sharp one has
narrow peak around the peak redshift, it has much large correlation
with the lensing potential in this narrow range.  The degree of the
correlation takes maximum value at certain redshift, and it gradually
decreases above the redshift.  This fact reflects that $\Delta f_{\rm
  NL}$ slowly increases with increasing of the peak redshift above the
certain redshift.  On the other hand, because the gradual distribution
model has a wide peak, the galaxy distribution gas the lower degree of
correlation with the lensing potential, even though the correlation
exists over the wider range in $k$-space.  In other words, to
constrain the primordial non-Gaussianity from galaxy-CMB lensing
cross-correlation one should select the galaxies whose correlation
with the CMB lensing potential becomes maximum.

\begin{table}[th]
\begin{ruledtabular}
\begin{tabular}{c|p{0.5pt}cp{0.5pt}cp{0.5pt}cccp{0.5pt}cccp{0.5pt}ccc}
&&  
\multicolumn{1}{c}{CMB} && 
\multicolumn{1}{c}{CMB} &&
\multicolumn{3}{c}{Case (i)} &&
\multicolumn{3}{c}{Case (ii)} &&
\multicolumn{3}{c}{Case (iii)} \\ 
\cline{3-3} \cline{5-5} \cline{7-9} \cline{11-13} \cline{15-17} \vspace{-3mm}\\ 
 && No lensing && ~Lensing~ &&
  $z_0 = 0.6$ & $z_0 = 1.2$ & $z_0 = 1.8$ &&
  $z_0 = 0.6$ & $z_0 = 1.2$ & $z_0 = 1.8$ &&
  $z_0 = 0.6$ & $z_0 = 1.2$ & $z_0 = 1.8$  \\ \hline \hline
$100 \Omega_b h^2$
&&  0.0243 &&  0.0225 &&  0.0225 &  0.0225 &  0.0225 &&  0.0224 &  0.0225 &  0.0225 &&  0.0223 &  0.0223 &  0.0224 \\
$\Omega_c h^2$
&&  0.00222 &&  0.00214 &&  0.00213 &  0.00214 &  0.00214 &&  0.00212 &  0.00215 &  0.00216 &&  0.00211 &  0.00213 &  0.00214 \\
$\Omega_{\Lambda}$
&&  0.1922 &&  0.0530 &&  0.0519 &  0.0527 &  0.0528 &&  0.0635 &  0.0677 &  0.0680 &&  0.0504 &  0.0526 &  0.0527 \\
$\tau$
&&  0.00553 &&  0.00457 &&  0.00457 &  0.00457 &  0.00457 &&  0.00463 &  0.00463 &  0.00463 &&  0.00456 &  0.00456 &  0.00456 \\
$f_\nu$
&&  0.0384 &&  0.0107 &&  0.0107 &  0.0106 &  0.0107 &&  0.0120 &  0.0120 &  0.0121 &&  0.0107 &  0.0106 &  0.0106 \\
$Y_{\rm He}$
&&  0.0159 &&  0.0152 &&  0.0152 &  0.0152 &  0.0152 &&  0.0153 &  0.0154 &  0.0154 &&  0.0151 &  0.0152 &  0.0152 \\
$n_s$
&&  0.01016 &&  0.00937 &&  0.00933 &  0.00934 &  0.00936 &&  0.00926 &  0.00932 &  0.00936 &&  0.00924 &  0.00929 &  0.00933 \\
$\Delta_{\rm R}^{2}(k_0)\times 10^9$
&&  0.0295 &&  0.0237 &&  0.0237 &  0.0237 &  0.0237 &&  0.0242 &  0.0243 &  0.0243 &&  0.0237 &  0.0237 &  0.0237 \\
$w$
&&  0.651 &&  0.197 &&  0.191 &  0.195 &  0.195 &&  0.253 &  0.269 &  0.270 &&  0.187 &  0.195 &  0.195 \\
$N_{\rm eff}$
&&  0.135 &&  0.116 &&  0.116 &  0.116 &  0.116 &&  0.116 &  0.117 &  0.117 &&  0.116 &  0.116 &  0.116 \\
$\alpha$
&&  0.00841 &&  0.00755 &&  0.00749 &  0.00752 &  0.00754 &&  0.00761 &  0.00767 &  0.00769 &&  0.00745 &  0.00751 &  0.00753 \\ \hline
$f_{\rm NL}$
&&    -----   &&    -----   &&105.7 & 39.9 & 27.0 &&104.9 & 38.3 & 25.9 &&100.8 & 36.9 & 25.1 \\
$b_0$
&&    -----   &&    -----   &&  0.0682 &  0.0718 &  0.0931 &&  0.0703 &  0.0826 &  0.1076 &&  0.0662 &  0.0707 &    0.0911 
\end{tabular}
\caption{Constraints on cosmological parameters from the Fisher matrix 
  methods for each case. We show the marginalized 1$\sigma$ errors for 
  the eleven-parameter or 
  thirteen-parameter models. We assume Planck for CMB survey. For galaxy survey 
  we assume that sky coverage and total number of galaxy sampling 
  are $f_{\rm sky} = 0.1$ and $N_{\rm g} = 10^6$.
  The parameters of galaxy sampling model are taken as 
  $\alpha = 0.5$, $\beta = 3.0$ for all cases. 
  \label{tb:ParError}}
\end{ruledtabular}
\end{table}

Finally, we focus on the observing redshift dependence for the
constraints on the primordial non-Gaussianit $f_{\rm NL}$.  We found
that the constraints on $f_{\rm NL}$ considerably depends on the mean
redshift of the observations ${z_{\rm m}}$, which is related with
model parameter $z_{0}$ by Eq. (\ref{meanredshift}). We show its
redshift dependence in Fig.~\ref{fig:ParErrorComp}. The errors of
$f_{\rm NL}$ rapidly decrease with redshift at low redshift, $z_0 \leq
1$, while it increases slowly at high redshift, $z_0 \geq 1$.  The
effect of the primordial non-Gaussianity through the scale-dependent
bias becomes large at high redshift, so that this is the reason why
the smaller error of $f_{\rm NL}$ can be obtained when the higher
redshift is probed.  On the other hand, the auto-correlation signal of
the galaxy distribution becomes small with increasing redshift. This
is an opposite effect to that from the scale-dependent bias for
constraint on $f_{\rm NL}$ and this is the reason why the error of
$f_{\rm NL}$ gradually increases at higher redshift, in particular, in
Case~(i) There are two reasons for increasing of the error of $f_{\rm
  NL} $ at high-redshift.  One is due to galaxy sampling model defined
by Eq.~(\ref{Gdist}). This analytic form may drop the information of
the low-redshift galaxies at high $z_0$. The other is that the
cross-correlation between lensing potential and galaxy becomes weak at
high redshift $z_0$, as seen in Fig.~\ref{fig:Clpg} for $f_{\rm NL} =
0$ (solid line).

\begin{figure}[thb]
\centering
\includegraphics[clip,keepaspectratio=true,width=0.5
\textwidth]{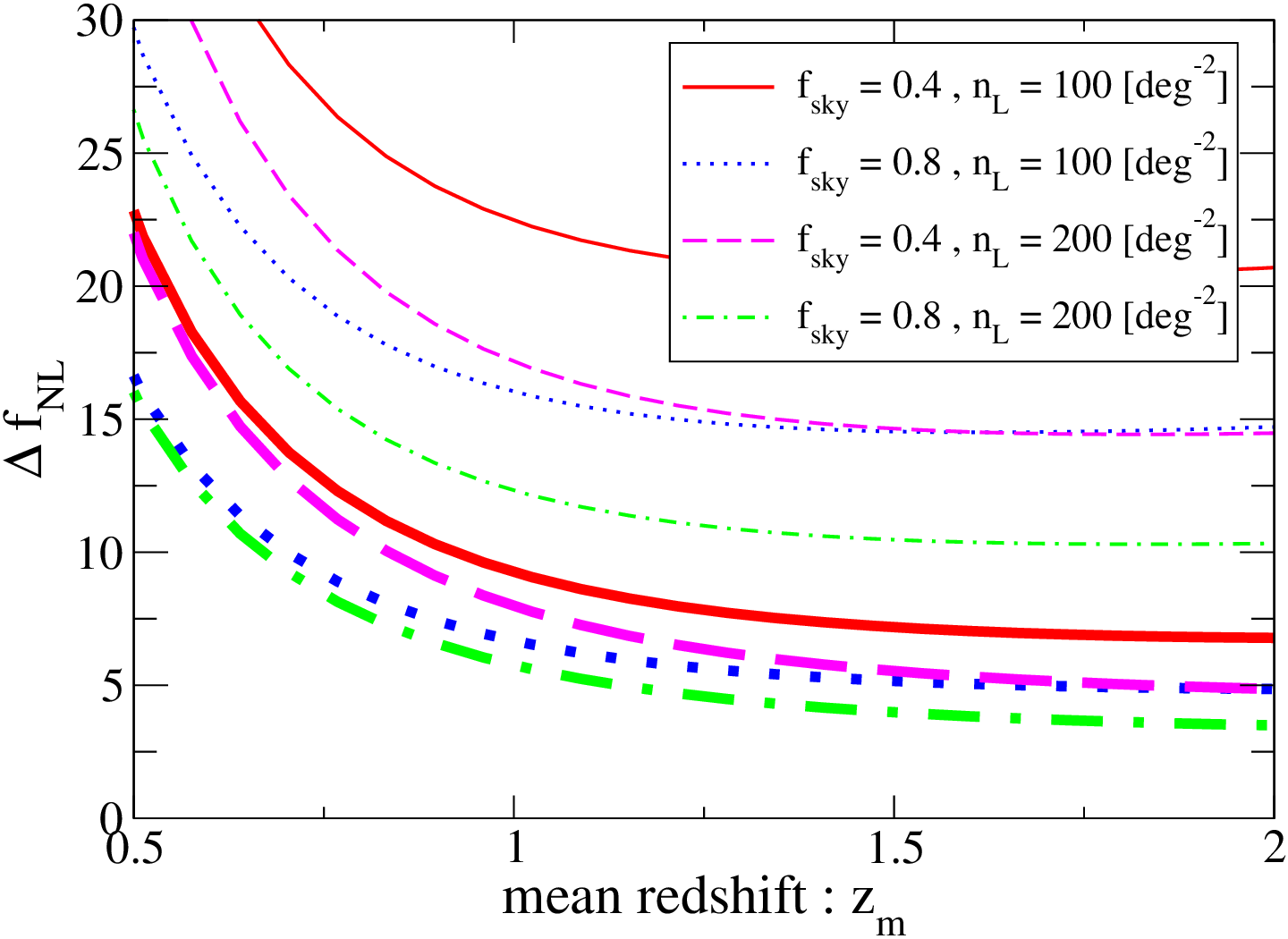}
\caption{ The 1$\sigma$ error of $f_{\rm NL}$ for wide galaxy survey
  with different mean redshift.  We assume Planck for CMB survey, and
  galaxy sampling model parameters are adopted as $\alpha = 0.5$,
  $\beta = 3.0$.  The thin and thick lines correspond to $b_0 = 2$ and
  $b_0 = 4$ cases, respectively.
\label{fig:dfNL-FS}}
\end{figure}

Moreover, we compare the constraints from various galaxy survey
conditions and linear bias parameters $b_0$ in Fig.~\ref{fig:dfNL-FS}.
From this figure, we clearly see that both "depth" and "width" for
galaxy survey are important for constraint on $f_{\rm NL}$
and what should be stressed is that the case of large bias $b_0$
constrains more strictly than the case of low bias. The objects with
large bias are affected by the primordial non-Gaussianity more
strongly than the objects with low bias, so one of the key points to
put constraint on the primordial non-Gaussianity is to explore the
highly biased objects. This results indicate that some galaxy survey
exploring the highly biased objects could constrain $f_{\rm NL}$ in
less than 10 even with Planck, 
for example, $\Delta f_{\rm NL} \sim ~ 5$ at $z_{\rm m} = 2.0$.

The estimations given above do not take into account conditions of
realistic observations, because we vary only the peak redshift $z_0$
fixing the survey area $f_{\rm sky}$ and the available galaxy samples
$N_{\rm g}$. In fact, in the real galaxy survey the survey area and
the available galaxy samples will also vary due to the change of the
observed mean redshift because the total observation time is finite as
explained in Sec.~\ref{sec:Modeling}.  Accounting the realistic galaxy
survey condition, what strategy should we develop for constraining the
primordial non-Gaussianity, $e.g.$ deep survey or wide survey ?  In
the next section, we search for the best condition for constraining
$f_{\rm NL}$.

\subsection{For the Galaxy Survey with Modeling}
\begin{figure}[t]
\centering
\includegraphics[clip,keepaspectratio=true,width=0.45\textwidth]{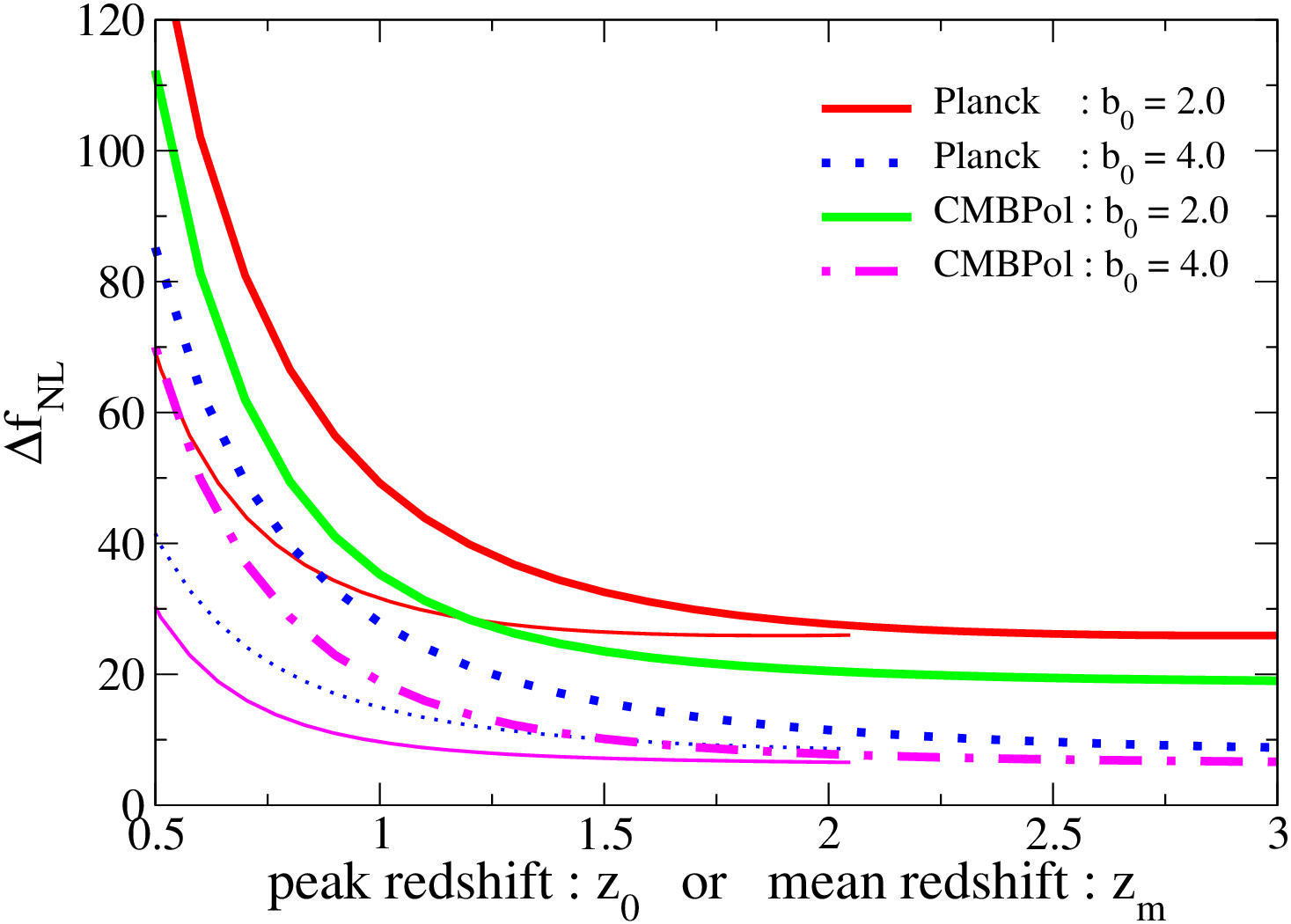}
\hspace{5mm}
\includegraphics[clip,keepaspectratio=true,width=0.45\textwidth]{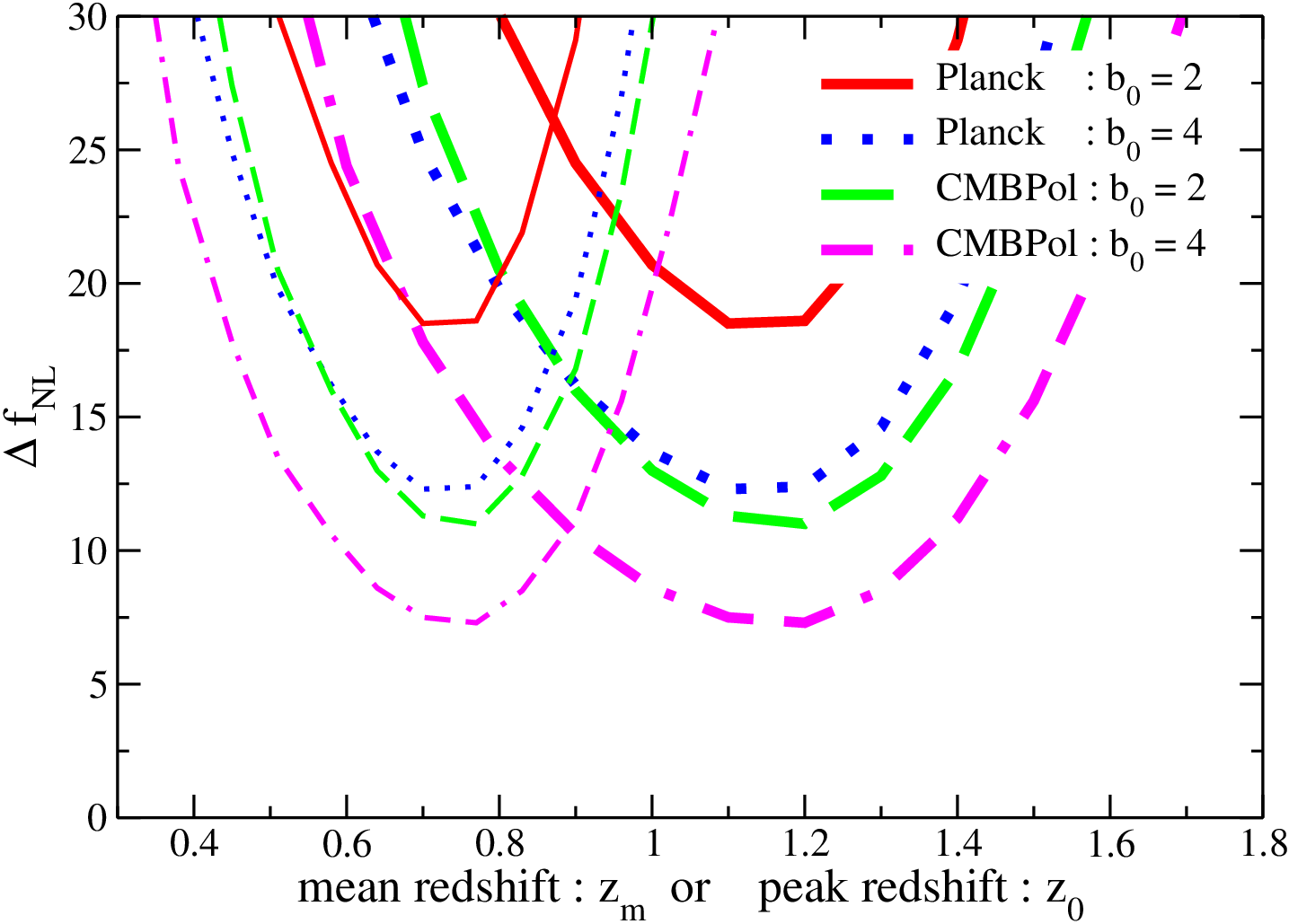}
\caption{
The 1$\sigma$ error of $f_{\rm NL}$ in the case with fixed sky coverage 
$f_{\rm sky} = 0.1$ and total number of galaxy $N_{\rm g} = 10^6$ (Left), and 
in the case with modeling for HSC survey (Right). For the CMB experiment 
we consider two cases. 
The thick lines show the errors against peak redshift $z_0$ and the thin lines 
are against mean redshift $z_{\rm m}$. 
\label{fig:ParErrorM}}
\end{figure}

Assuming observations like HSC, we estimate a more realistic
constraint on $f_{\rm NL}$ using the survey model introduced in
Sec.\ref{sec:Modeling}. We show the results in Fig.\ref{fig:ParErrorM}
and the relations between various survey parameters, namely, peak
redshift $z_0$, survey area $f_{\rm sky}$, number density of sampling
galaxy $n_{\rm L}$ and exposure time $t_{\rm exp}$ with a fixed observation
time in Table~\ref{tb:ParRelation}.

In Fig.~\ref{fig:ParErrorM}, we find that there is a minimal point for
the error of $f_{\rm NL}$ around ${z_{\rm m}} \simeq 0.7$ and in this
point, the constraints on $f_{\rm NL}$ are $\Delta f_{\rm NL} \sim
~20$ for $b_0 = 2$ and $\Delta f_{\rm NL} \sim ~10$ for $b_0 = 4$.
From this result, the target redshift we should observe is not deep
enough, in which case the surface number density of the sample
galaxies is relatively small, and the observation area can be wide,
$n_{\rm L} \simeq 7.0$ $[{\rm arcmin}^{-2}]$ and $f_{\rm sky} \simeq
0.35$ (at $z_{\rm m} = 0.7$). (see Table~\ref{tb:ParRelation}.)  For
the question whether we should make wide or deep survey, the answer is
that we should select the wide survey. The signal of the primordial
non-Gaussianity in the scale-dependent bias $b (z, k)$ is more
significant in relatively large-scale regions, as seen in
Fig.~\ref{fig:Clpg}.  The noise contributions in large-scale regions
are dominant by the cosmic variance due to the finiteness of survey
area.  On the other hand, small scale region is dominated by shot
noise related to the surface number density of the galaxy samples
$n_{\rm L}$, although the signal of the primordial non-Gaussianity is
not sensitive there. Therefore, we conclude that the better
constraining the primordial non-Gaussianity, $f_{\rm NL}$, prefers
wide area survey to the deep survey.  However, note that this
conclusion is derived by assuming Planck and HSC experiments with a
fixed observation time.  Accounting for the redshift dependence of the
effect of the primordial non-Gaussianity through scale-dependent bias,
we should keep it in mind that the deep survey also becomes important
to put a tighter constraint on $f_{\rm NL}$.

\section{Summary and Discussion}
\label{sec:Summary}
In this paper we have estimated the constraints on the cosmological
parameters newly taking into account the cross-correlation between CMB
lensing and galaxy angular distributions. In particular, we have
focused on the constraint on the primordial non-Gaussianity through
the scale-dependent bias $b (z, k)$ and estimated how much the
cross-correlation between CMB lensing and galaxy would improve the
constraint by the Fisher matrix analysis. 
In order to make the most general Fisher matrix analysis with CMB and
galaxy survey experiments, we have taken into account the all the
auto- and cross-correlations available which is expressed by the
8$\times$8 covariance matrix as Eq. (\ref{eq:Cov}). 
We have paid particular attention to the CMB and galaxy survey just
coming up now, namely Planck satellite and HSC, 
and also to the future experiments like CMBPol,
LSST and so on. 
Our estimations are mainly based on the Planck and HSC
surveys, however we also show some cases for comparison 
when CMBPol or ambitious survey conditions are assumed.

As for the constraints on the conventional cosmological parameters,
the improvement can not be expected very much from the simple estimate
in which the Fisher matrices for the CMB and galaxy surveys are
combined properly even if the cross-correlations are properly taken
into account.  However, focusing on the determination of $f_{\rm NL}$,
we found that the role of the cross-correlation between CMB and galaxy
is important, especially the one between CMB lensing potential and
galaxy $\cpg$ contributes the determination of $f_{\rm NL}$.

We have estimated the constraints on the $f_{\rm NL}$ for the coming
experiments.  First, we gave the rough estimation in cases where the
survey area $f_{\rm sky}$ and galaxy samples $N_{\rm g}$ are fixed and only
peak redshift $z_0$ is varied. It was found that the keys for more strictly 
constraining the primordial non-Gaussianity are
observing higher redshift and larger biased objects. 
Second, considering the realistic observations with fixed observation
time, we have estimated the constraint on $f_{\rm NL}$, adapting the
galaxy survey model in Ref.~\cite{Yamamoto-07}, which is scaled to the
HSC survey. 
As a result we found on optimized target redshift to be $z_0 \simeq
1$, which brings us to a conclusion that we should make a wide survey
rather than a deep survey for constraining the primordial non-Gaussianity with
HSC-like observation. 
This is because the effect of the primordial non-Gaussianity through
the scale-dependent bias is significant on large scales rather than
small scales. The large-scale region is dominated by the
cosmic variance related to the survey area while
the small-scale region is dominated by the shot noise term related to
the number of the galaxy samples. 
Therefore, we should explore the wide survey area rather than
observing a lot of galaxies. 
However, we should keep it in mind that high-redshift and highly-biased objects 
are much affected by primordial non-Gaussianity, so that
the deep survey will be also essential, in the future.

The constraints on the primordial non-Gaussianity expected from HSC
survey with Planck are : $\Delta f_{\rm NL} \sim ~20$ for $b_0 = 2$
and $\Delta f_{\rm NL} \sim ~10$ for $b_0 = 4$.  Slosar et
al.~\cite{Slosar-08} obtained constraints on $f_{\rm NL}$ for highly
biased tracers using available luminous red galaxy (LRG) or quasar
(QSO) data from Sloan Digital Sky Survey (SDSS) \cite{Ho-08} and CMB
data from WMAP 5, with the error of $\Delta f_{\rm NL} \sim 30$.  The
reason why the combination of HSC and Planck observations does not
make significant improvement over the current constraints is explained
as below.  We have considered only 800 hours for HSC and normal
galaxies. The SDSS data is obtained over longer period of time, and
QSOs seem to be observed more than normal galaxies at high-redshift.
These constraints are weaker than those expected with the CMB
bispectrum constraints achievable with an ideal CMB experiment,
$\Delta f_{\rm NL} \sim ~{\rm few}$ (Ref.~\cite{Yadav-07,
  Liguori-08}).  However, the constraint on $f_{\rm NL}$ presented in
this paper depends highly on the galaxy survey condition, $i.e.,$
survey area $f_{\rm sky}$, number density of sample galaxies $n_{\rm
  L}$ and observing peak redshift $z_0$.  It is found that with some
galaxy survey with Planck $f_{\rm NL}$ could achieve $\Delta f_{\rm
  NL} \sim ~ 5$ at ${z_{\rm m}} = 2.0$ (Fig.~\ref{fig:dfNL-FS}).
Recently it is reported that an ambitious future galaxy survey (like
the LSST survey), which provides large survey area of 30,000 ${\rm
  deg^2}$ and highly biased galaxy samples, can measure the primordial
non-Gaussianity with the order $\Delta f_{\rm NL} \sim ~2-5$
\cite{Carbone-10}.  The other method using the full covariance of
cluster counts for Dark Energy Survey (DES) can yield $\Delta f_{\rm
  NL} \sim 1-5$ \cite{Cunha-10}.  In Ref.~\cite{Carbone-10, Cunha-10},
they add the Planck and galaxy survey Fisher matrices, $i.e.,$ $F_{ij}
= F_{ij}^{\rm galaxy} + F_{ij}^{\rm CMB} $, and do not include the
cross-correlation between CMB and galaxy survey, $i.e.,$ $\ctg$ an
$\cpg$, so that more tight constraint on $f_{\rm NL}$ may be expected
with these cross-correlations.  In any case, it is worth pursuing how
well we can put a constraint on non-Gaussianity of the local-type from
the large-scale structure because it contains information on
non-Gaussianity at different epoch from CMB and thus the constraint
through the scale-dependent bias will be an important cross check
against the CMB bispectrum.

\begin{table}[th]
\begin{ruledtabular}
  \begin{tabular}{ccccc}
    $z_0$ & $z_{\rm m}$ & $\Delta f_{\rm NL}$ & $n_{\rm L} {\rm [arcmin^{-2}]}$ & $f_{\rm sky}$\\
    \hline
    0.8 & 0.51 & 30.1 &  0.9 & 0.42 \\
    0.9 & 0.58 & 24.5 &  1.9 & 0.41 \\
    1.0 & 0.64 & 20.7 &  3.7 & 0.40 \\
    1.1 & 0.70 & 18.5 &  7.0 & 0.35 \\
    1.2 & 0.77 & 18.6 & 12.4 & 0.25  \\
    1.3 & 0.83 & 21.9 & 20.9 & 0.14 \\
    1.4 & 0.90 & 29.1 & 34.0 & 0.06 \\
    1.5 & 0.96 & 40.5 & 53.5 & 0.02 
\end{tabular}
\caption{ The relation of parameters with fixed observation time,
  between the peak redshift $z_0$, the mean redshift $z_{\rm m}$, the
  $1\sigma$ error of $f_{\rm NL}$, the number density of sampling
  galaxy $n_{\rm L}$ and the survey area $f_{\rm sky}$, for the
  HSC-like survey ($b_0=2$). \label{tb:ParRelation}}
\end{ruledtabular}
\end{table}

\acknowledgments{ We thank A.Taruya, T.Namikawa, S.Saito, M. Alvarez,
  C.Cunha, D.Huterer and O.Dor\'{e} for useful discussion and
  comments.  We acknowledge support from JSPS (Japan Society for
  Promotion of Science) Grant-in-Aid for Scientific Research
  No. 21740177, 22012004 (KI), Grant-in-Aid for Nagoya University
  Global COE Program "Quest for Fundamental Principles in the
  Universe: from Particles to the Solar System and the Cosmos", from
  the Ministry of Education, Cluster, Sports, Science, and Technology,
  Grant-in-Aid for Scientific Research (C), 21540263, 2009 (TM), and
  Grant-in-Aid for Scientific Research on Priority Areas No. 467
  "Probing the Dark Energy through an Extremely Wide and Deep Survey
  with Subaru Telescope." This work is supported in part by JSPS
  Core-to-Core Program "International Research Network for Dark
  Energy."  }

\bibliography{CMB-Galaxy}


\end{document}